\documentclass[iop]{emulateapj}

\usepackage{comment}
\usepackage{amsmath}
\usepackage{graphicx,color}
\usepackage{hyperref}
\usepackage{indentfirst}
\usepackage{verbatim}
\usepackage{morefloats}
\usepackage{natbib}
\usepackage{epstopdf}
\usepackage{appendix}
\shorttitle{Line De-confusion in Intensity Mapping}
\shortauthors{Cheng et al.}

\begin{document}

\title{Spectral Line De-confusion in an Intensity Mapping Survey}

\author{Yun-Ting Cheng$^{1,2}$, Tzu-Ching Chang$^2$, James Bock$^{1,3}$, C. Matt Bradford$^{3,1}$, Asantha Cooray$^{4}$}
\address{$^1$California Institute of Technology, 1200 E. California Blvd, Pasadena, CA 91125, U.S.A.}
\address{$^2$Institute of Astronomy and Astrophysics, Academia Sinica, 1 Roosevelt Rd, Section 4, Taipei, 10617, Taiwan}
\address{$^3$NASA Jet Propulsion Laboratory, 4800 Oak Grove Drive, Pasadena, CA 91109, U.S.A.}
\address{$^4$Department of Physics and Astronomy, University of California, Irvine, CA 92697, U.S.A.}

\begin{abstract}
Spectral line intensity mapping has been proposed as a promising tool to efficiently probe the cosmic reionization and the large-scale structure.  Without detecting individual sources, line intensity mapping makes use of all available photons and measures the integrated light in the source confusion limit, to efficiently map the three-dimensional matter distribution on large scales as traced by a given emission line.  One particular challenge is the separation of desired signals from astrophysical continuum foregrounds and line interlopers.  Here we present a technique to extract large-scale structure information traced by emission lines from different redshifts, embedded in a three-dimensional intensity mapping data cube. The line redshifts are distinguished by the anisotropic shape of the power spectra when projected onto a common coordinate frame.  We consider the case where high-redshift [CII] lines are confused with multiple low-redshift CO rotational lines.  We present a semi-analytic model for [CII] and CO line estimates based on the cosmic infrared background measurements, and show that with a modest instrumental noise level and survey geometry, the large-scale [CII] and CO power spectrum amplitudes can be successfully extracted from a confusion-limited data set, without external information. We discuss the implications and limits of this technique for possible line intensity mapping experiments.
\end{abstract}


\section{Introduction}
Line intensity mapping has emerged as a promising tool to probe the three-dimensional structure of the Universe. Several emission lines have been proposed to uniquely trace the cosmic reionization process, revealing properties of the ionizing sources and the intergalactic medium at high redshifts, and to efficiently map the large-scale matter distribution in a large cosmic volume, suitable for cosmological studies at lower redshifts.  

In contrast to traditional large-scale structure surveys, intensity mapping (IM) operates in the confusion limited regime without thresholding to identify individual sources;  rather, IM makes use of integrated light emission from all sources, including unresolved faint galaxies, to statistically measure properties of the light tracers and the underlying matter distribution.  In addition, with line intensity mapping (LIM), where the tracer is a particular spectral line emission, the three-dimensional matter distribution can be faithfully represented on large cosmic scales. The 21cm hyperfine emission from neutral hydrogen \citep{1990MNRAS.247..510S,1997ApJ...475..429M,2008MNRAS.383..606W,2008PhRvL.100i1303C}, the CO rotational lines \citep{2008A&A...489..489R,2010JCAP...11..016V,2011ApJ...730L..30C,2011ApJ...741...70L,
2011ApJ...728L..46G,2014MNRAS.443.3506B, 2013ApJ...768...15P, 2015JCAP...11..028M,2015ApJ...814..140K,2016ApJ...817..169L}, the [CII] 157.7 $\mu$m fine structure line \citep{2012ApJ...745...49G,2014ApJ...793..116U,2015ApJ...806..209S,2015MNRAS.450.3829Y}, and the Lyman-$\alpha$ emission line \citep{2013ApJ...763..132S,2014ApJ...785...72G,2014ApJ...786..111P,
2016MNRAS.455..725C} are amongst the most studied such tracers in the LIM regime. 

One of the main challenges in a line intensity mapping experiment is the separation of signals from the astrophysical foreground continuum emissions and line interlopers.  The continuum foreground issue has been studied mostly in the context of 21cm intensity mapping, where the Galactic and extragalactic synchrotron and free-free radiations overwhelm the expected signals by several orders of magnitude \citep[e.g.,][]{2006PhR...433..181F,2006ApJ...648..767M,2009ApJ...695..183B,
2012MNRAS.419.3491L,2012ApJ...756..165P,2012MNRAS.423.2518C,2015ApJ...815...51S}; line interlopers, on the other hand, are a pressing issue for other line probes in the electromagnetic spectrum that is crowded with other line features. 
  
Several studies have proposed strategies for deblending lines in an intensity mapping survey, by means of masking and cross-correlation. The masking technique makes use of an external galaxy catalog from galaxy surveys to identify and remove bright sources in order to reduce potential foreground contaminations \citep{2015MNRAS.452.3408B,2015MNRAS.450.3829Y,2015ApJ...806..209S}. On the other hand, cross-correlation of a LIM survey with an external data set tracing the same cosmic volume can help extract signals of interest;  the method has been proposed in particular in the studies of reionization\citep{2009ApJ...690..252L,2010JCAP...11..016V,2012ApJ...745...49G, 2014ApJ...785...72G,2015aska.confE...4C,2015ApJ...806..209S}, and has been successfully applied to extract LIM signals at lower redshifts against continuum foregrounds \citep{2010Natur.466..463C,2013ApJ...763L..20M, 2016MNRAS.457.3541C}. Aside from these two methods, \citet{2015ApJ...806..234K} makes use of the companion lines to directly identify [CII] line intensity in each voxel. \citet{2014arXiv1403.3727D} propose to use angular fluctuations of the light to reconstruct the 3D source luminosity density.

In an intensity mapping experiment, the intrinsic observing coordinates are angular and spectral coordinates defined by the instrument and survey geometry, which, given a known redshift, are mapped into comoving coordinates in the redshift space, before a power spectrum is computed.  In the linear regime, any line tracers supposedly follow the matter distribution, which is isotropic in their respective real-space coordinates.  If, however, without a priori knowledge of redshifts, lines at different redshifts embedded in an observing volume are blindly projected into the same comoving coordinates, they will exhibit different anisotropic 3D shapes in that frame, due to the incorrect redshift projection. This is key to the line separation technique we employ in this paper.  The idea has been previously suggested by \citet{2010JCAP...11..016V} and \citet{2014ApJ...785...72G}, and recently investigated by \citet{2016arXiv160405737L}.

To demonstrate this technique, we consider a 3D LIM observing volume with high-redshift [CII] emissions blended with multiple lower-redshift CO rotational lines. We present a halo-model based formalism to estimate [CII] and CO line strengths and power spectra across redshifts.  After projecting the observed volume onto a common comoving frame, the resulting total power spectrum is a superposition of [CII] and CO power spectra at different redshifts, each with a different but predictable 3D shape due to the projection which we use as templates.  We generate simulated data and use the Markov Chain Monte Carlo (MCMC) formalism to extract power spectrum parameters based on the templates.

The paper is organized as follows: In Sec.~\ref{S:model}, we describe a model to estimate the [CII] and CO power spectra across redshifts, and the formalism for expressing the 3D power spectra of both lines in the comoving frame of [CII]. In Sec.~\ref{S:MCMCinfer}, we discuss the details of the template fitting and MCMC implementation. The results are presented in Sec.~\ref{S:results}. In Sec.~\ref{S:discussion}, we discuss the implication and limitation of our method, and conclude in Sec.~\ref{S:conclusion}. Throughout this paper, we consider a flat $\Lambda$CDM cosmology with $n_s=0.97$, $\sigma_8=0.82$, $\Omega_m=0.26$, $\Omega_b=0.049$, $\Omega_\Lambda=0.69$, and $h=0.68$, consistent with the latest measurement from \textit{Planck} \citep{2015arXiv150201589P}.

\section{Power Spectrum Modeling}\label{S:model}
Here we provide an estimate of the [CII] and CO power spectra as a function of redshift. Our modeling is based on the halo model formalism \citep{2002PhR...372....1C}.  With the [CII] and CO luminosity ($L_{CII},\ L_{CO}$) and halo mass ($M$) relations, the power spectrum of a spectral line intensity field can be calculated with the halo model.

We build our model (referred to as the CIB model hereafter) based on \citet{2014A&A...571A..30P}; the authors fitted the $L_{IR}$ and and halo mass relation with cosmic infrared background (CIB) emission as measured by \textit{Planck}. Details of the CIB model are provided in Appendix~\ref{A:M_SFR}.

Then we connect the $L_{IR}$ to  $L_{CII}$ and $L_{CO}$. We adopt these relations based on observations and simulations in the literature.  See Appendix~\ref{A:LIR_Lline} for more details.

\subsection{Power Spectrum With Halo Model}\label{S:PS_halo}
With the luminosity - halo mass relation at hand, we derive the power spectrum in the halo model framework. The comoving power spectrum consists of the one-halo and two-halo terms, which account for the correlation within the halos and between halos, respectively. The one-halo term of the [CII] or CO line is given by
\begin{align}
P^{line}_{1h}(k,z)&=\int_{M_{min}}^{M_{max}} dM \frac{dN}{dM}\left | u(k|M) \right |^2\nonumber\\
&\times\left (  \frac{L_{line}(M,z)}{4\pi D_L^2(z)}y_{line}(z)D_A^2(z)\right )^2,
\end{align}
where $L_{line}(M,z)$ is the luminosity of [CII] or CO for a given halo mass $M$ at redshift $z$, $D_L$ is the proper luminosity distance, $D_A$ is the comoving angular diameter distance, $u(k|M)$ is the Fourier transform of the normalized halo density profile, and we adopt the NFW profile in this work \citep{1996ApJ...462..563N}. $y_{line}(z)\equiv d\chi/d\nu=\lambda_{line}(1+z)^2/H(z)$, where $\chi$ is the comoving distance, $H$ the Hubble parameter, $\nu$ the observed frequency, and $\lambda_{line}$ the rest frame wavelength of the line.

The two-halo term can be written as
\begin{align}
&P^{line}_{2h}(k,z)=[ \int_{M_{min}}^{M_{max}} dM \frac{dN}{dM}\left | u(k|M) \right |b(M,z)\nonumber \\ 
&\times\left (  \frac{L_{line}(M,z)}{4\pi D_L^2(z)}y_{line}(z)D_A^2(z)\right )]^2P_{lin}(k),
\end{align}
where $b(M,z)$ is the halo bias \citep{1999MNRAS.308..119S}, and $P_{lin}(k)$ is the linear matter power spectrum.

We also consider the shot noise power spectrum due to the discretization of sources:
\begin{equation}\label{E:Psh}
P^{line}_{sh}(z)=\int_{M_{min}}^{M_{max}} dM \frac{dN}{dM}\left (  \frac{L_{line}(M,z)}{4\pi D_L^2(z)}y_{line}(z)D_A^2(z)\right )^2.
\end{equation}

The total comoving power spectrum is thus
\begin{equation}
P^{line}_{tot}(k,z)=P^{line}_{1h}(k,z)+P^{line}_{2h}(k,z)+P^{line}_{sh}(z).
\end{equation}

\begin{figure}
\begin{center}
\includegraphics[width=\linewidth]{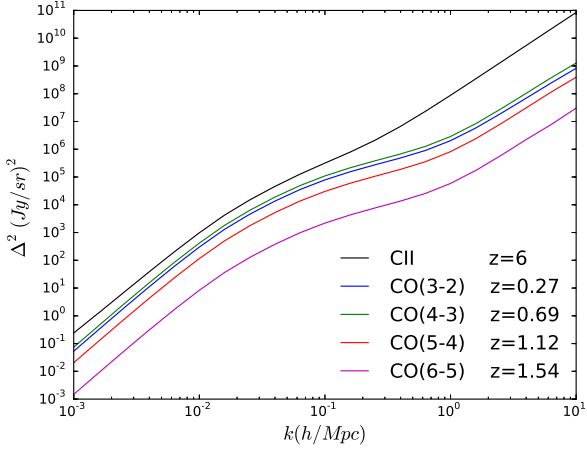}
\caption{\label{F:1DPS} The comoving isotropic power spectra of [CII] at $z=6$ and CO interlopers at lower redshfits in the same observing frequency range. }
\end{center}
\end{figure}

Fig.~\ref{F:1DPS} shows the comoving isotropic power spectrum of [CII] at $z=6$ based on the CIB model, and the lower-redshift CO power spectra which overlap in the same observing frequency range. 

\subsection{3D Power Spectrum}
The [CII] and CO power spectra $P(\textbf{k})$ is isotropic in real-space on large scales, so the power spectra only depend on $\left |  \textbf{k}\right |$ but not the direction of the $\textbf{k}$ vector. However, an anisotropic feature comes in due to observational effects, which introduce the dependence on $\mu$, the cosine of the angle between the $\textbf{k}$ vector and the line-of-sight direction. Below we discuss the projection effect and the redshift space distortions (RSD) that give rise to the anisotropy in the 3D power spectrum.

\subsubsection{Projection Effect}\label{S:projection}
In a redshifted [CII] intensity mapping experiment, a CO line signal from redshift $z_{CO}$ will blend with the [CII] signal from $z_{CII}$ if they have the same observing frequency. The redshifts of these two lines follow 
\begin{equation}\label{E:nu_convert}
\nu_{obs}=\frac{\nu_{CII}}{(1+z_{CII})}=\frac{\nu_{CO}}{(1+z_{CO})},
\end{equation}
where $\nu_{obs}$ is the observed frequency, and $\nu_{CII}$ and $\nu_{CO}$ are the rest-frame frequencies of the [CII] and CO lines, which are 1902 GHz and 115$J$ GHz for the $J$ to $J-1$ transition, respectively.

Both [CII] and CO power spectra are isotropic in their respective comoving frames. However, in the confusion limit, we may incorrectly project the observed [CII] and CO signals onto the comoving frame of [CII] for power spectrum calculation. In this case, the CO power spectrum is no longer isotropic. This is caused by the different redshift projection factors parallel and perpendicular to the line-of-sight, which makes the CO 3D data cube stretched more in one direction than the other.  Below we provide the formalism for calculating the 3D power spectrum of a low-z CO signal projected onto the high-z [CII] comoving frame.

The projection in the direction parallel to the line-of-sight can be derived by considering an observed frequency range $d\nu_{obs}$, which corresponds to either the [CII] from a comoving size $R^\parallel_{CII}$ at $z_{CII}$, or the CO signal with comoving size $ R^\parallel_{CO}$ at $z_{CO}$. The relation between $R^\parallel_{CII}$ and $R^\parallel_{CO}$ is 
\begin{align}
\frac{R^\parallel_{CO}}{R^\parallel_{CII}}&=\frac{d\chi(z_{CO})/d\nu_{obs}}{d\chi(z_{CII})/d\nu_{obs}}=\frac{y_{CO}(z_{CO})}{y_{CII}(z_{CII})}\nonumber\\
&=\frac{\lambda_{CO}(1+z_{CO})^2/H(z_{CO})}{\lambda_{CII}(1+z_{CII})^2/H(z_{CII})}\nonumber \\
&=\frac{(1+z_{CO})/H(z_{CO})}{(1+z_{CII})/H(z_{CII})}.
\end{align}

Since Fourier wavenumber $k^\parallel \propto 1/R^\parallel$, we obtain
\begin{equation}
k^\parallel_{CO}=k^\parallel_{CII}\frac{H(z_{CO})(1+z_{CII})}{H(z_{CII})(1+z_{CO})} \equiv k^\parallel_{CII}r^\parallel(z_{CII},J),
\end{equation}
where $J$ labels the CO transition from $J$ to $J-1$, and $r^\parallel(z_{CII},J)$ is the conversion factor for projecting the scale at $z_{CO}$ to $z_{CII}$ in the parallel direction. 

The transverse scale of CO will be projected to the scale of [CII] corresponding to the same observed angle $\theta$. Hence, the scale conversion relation in the perpendicular direction is 
\begin{equation}
\theta =\frac{R_{CO}^\perp}{D_A(z_{CO})}=\frac{R_{CII}^\perp}{D_A(z_{CII})},
\end{equation} 
where $D_A$ is the comoving angular diameter distance. Thus
\begin{equation}
k^\perp_{CO}=k_{CII}^\perp \frac{D_A(z_{CII})}{D_A(z_{CO})}\equiv k_{CII}^\perp r^\perp(z_{CII},J),
\end{equation}

where $r^\perp(z_{CII},J)$ is the conversion factor for projecting the scale at $z_{CO}$ to $z_{CII}$ in the perpendicular direction.  Besides the shift in $k$ value, the projection changes the comoving voxel volume $V_{vox}$ and induces an amplitude change in the power spectrum. Since the power spectrum is proportional to $1/V_{vox}$ at fixed intensity fluctuation, the projected CO power spectrum needs to be multiplied by the change in voxel volume $r^\parallel (r^\perp)^2$.

The CO projected power spectrum $P^{prj}_{CO}$ can thus be written as
\begin{equation}
P^{prj}_{CO}(k^\perp_{CII},k^\parallel_{CII},z_{CII},J)=r^\parallel (r^\perp)^2P_{CO}(k_{CO},z_{CO}),
\end{equation}
where $k_{CO}=\sqrt{(k^\perp_{CO})^2+(k^\parallel_{CO})^2}$, and $P_{CO}$ is the comoving CO power spectrum. 

For completeness, we also write down the [CII] power spectrum in the same coordinate,
\begin{equation}
P^{prj}_{CII}(k^\perp_{CII},k^\parallel_{CII},z_{CII})=P_{CII}(k_{CII},z_{CII}),
\end{equation}
where $k_{CII}=\sqrt{(k^\perp_{CII})^2+(k^\parallel_{CII})^2}$.

\subsubsection{Redshift Space Distortions}\label{S:RSD}

Here we incorporate the redshift space distortion (RSD) effects.  We consider the linear Kaiser effect \citep{1987MNRAS.227....1K} describing the coherent motion of structure growth on large scales which enhances the power spectrum, and the suppression on small scales due to the non-linear virial motion, which we write as an exponential damping term \citep{1992LNP...408....1P}. The comoving one-halo and two-halo power spectrum can be written as \citep{2001MNRAS.321....1W}
\begin{align}\label{E:rsd_1h}
&P^{line}_{1h}(k,\mu,z)=(1+\beta\mu^2)^2 \nonumber\\
&\times\int_{M_{min}}^{M_{max}} dM \frac{dN}{dM}[\left (  \frac{L_{line}(M,z)}{4\pi D_L^2(z)}y_{line}(z)D_A^2(z)\right )^2\nonumber \\
&\times\left | u(k|M) \right |^2e^{-(k\sigma_v\mu)^2/2}],
\end{align}
\begin{align}\label{E:rsd_2h}
&P^{line}_{2h}(k,\mu,z)=P_{lin}(k)(1+\beta\mu^2)^2\nonumber\\
&\times\{ \int_{M_{min}}^{M_{max}} dM \frac{dN}{dM}[\left (  \frac{L_{line}(M,z)}{4\pi D_L^2(z)}y_{line}(z)D_A^2(z)\right ) \nonumber\\
&\times \left | u(k|M) \right |b(M,z)e^{-(k\sigma_v\mu)^2/2}]\}^2,
\end{align}
where $(1+\beta\mu^2)^2$ is the Kaiser effect and $e^{-(k\sigma_v\mu)^2/2}$ is the exponential damping term. $\beta\equiv f/\bar{b}_{line}$, where f=dlnD/dlna is the logarithm derivative of the linear growth rate $D(z)$ with respect to the scale factor $a=1/(1+z)$, and $\bar{b}_{line}$ is the luminosity-weighted bias of the tracer, which we consider to be a constant on large scales. $\sigma_v$ is the 1D velocity dispersion within halo mass $M$. Assuming the halos are isothermal, the velocity dispersion can be estimated as
\begin{equation}
\sigma_v^2=\frac{GM}{2r_{vir}},
\end{equation}
where $r_{vir}$ is the virial radius of the halo.
\\

Combining the projection and RSD effects, the projected CO power spectrum can be written as
\begin{align}\label{E:rsd_1h_co}
&P_{CO(1h)}^{prj}(k_{CII},\mu_{CII})=r^\parallel (r^\perp)^2(1+\frac{f(z_{CO})}{\bar{b}_{CO}}\mu_{CO}^2)^2 \nonumber\\
&\times\int_{M_{min}}^{M_{max}} dM \frac{dN}{dM}[\left (  \frac{L_{line}(M,z_{CO})}{4\pi D_L^2(z_{CO})}y_{CO}(z)D_A^2(z_{CO})\right )^2\nonumber\\
&\times\left | u(k_{CO}|M) \right |^2e^{-(k_{CO}\sigma_v\mu_{CO})^2/2}],
\end{align}
\begin{align}\label{E:rsd_2h_co}
&P_{CO(2h)}^{prj}(k_{CII},\mu_{CII})=r^\parallel (r^\perp)^2P_{lin}(k_{CO})(1+\frac{f(z_{CO})}{\bar{b}_{CO}}\mu_{CO}^2)^2  \nonumber\\
&\times\{[ \int_{M_{min}}^{M_{max}} dM \frac{dN}{dM}[\left (  \frac{L_{line}(M,z_{CO})}{4\pi D_L^2(z_{CO})}y_{CO}(z_{CO})D_A^2(z_{CO})\right )\nonumber\\
&\times  \left | u(k_{CO}|M) \right |b(M,z_{CO})e^{-(k_{CO}\sigma_v\mu_{CO})^2/2}]\}^2.
\end{align}
We define $\mu_{CII}$ and $\mu_{CO}$ to be the cosine angle of the $\textbf{k}_{CII}$ and $\textbf{k}_{CO}$ vectors with respect to the line-of-sight direction, respectively. For the [CII] power spectrum in its own comoving frame, only the RSD effect needs to be considered, so the [CII] 2D power spectrum has the form given by Eq.~(\ref{E:rsd_1h}) and Eq.~(\ref{E:rsd_2h}).

\begin{figure}
  \includegraphics[width=\linewidth]{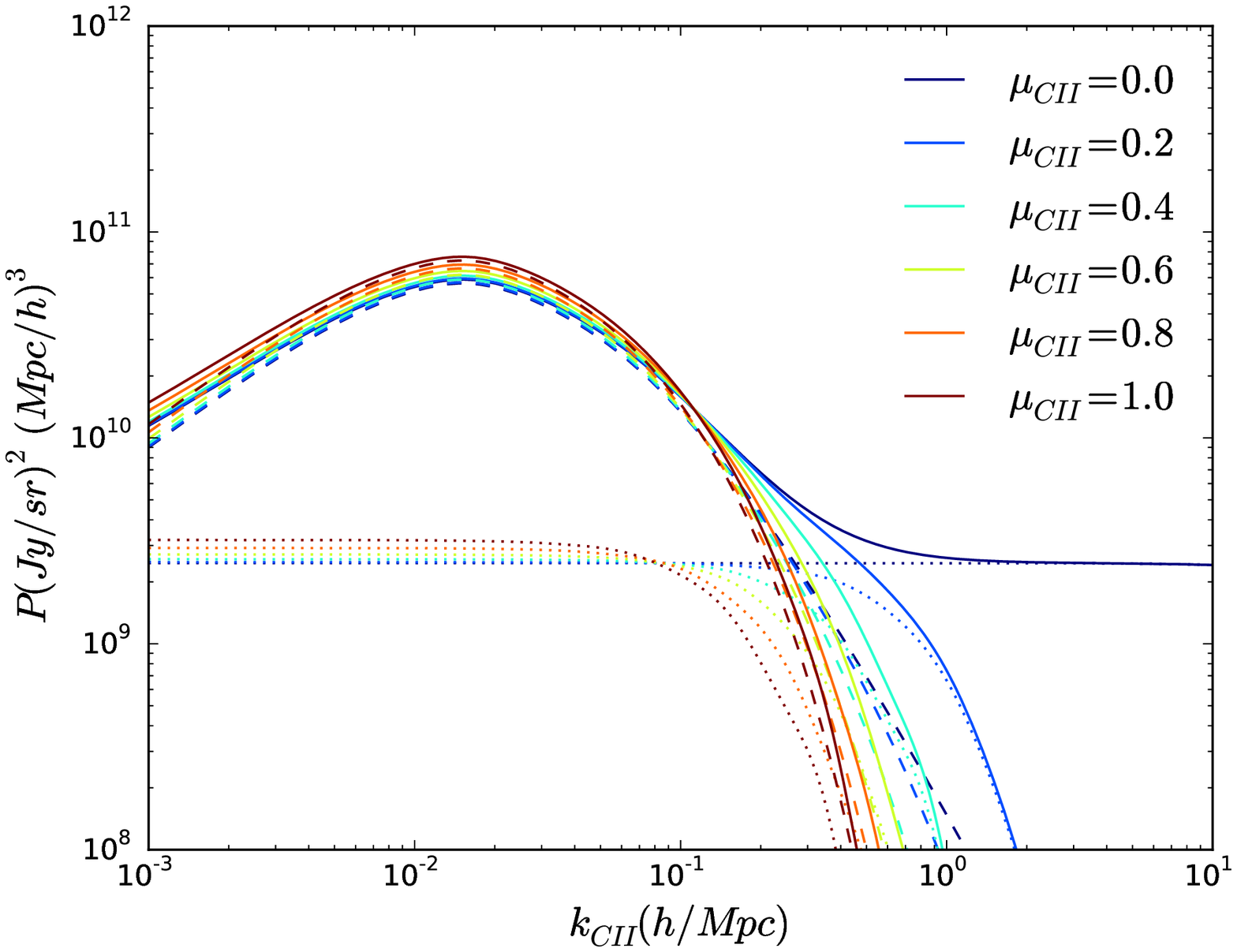}
  \includegraphics[width=\linewidth]{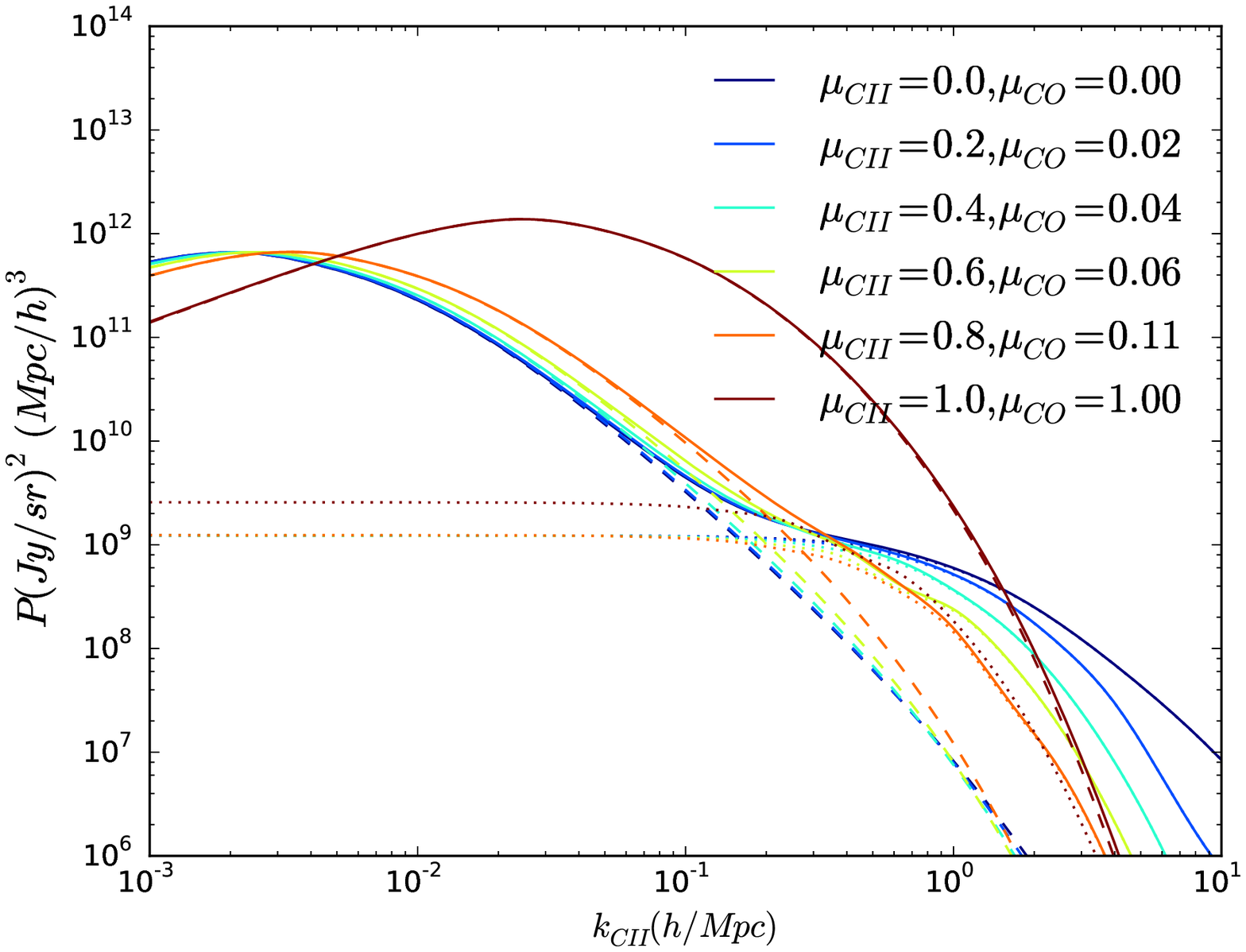}
  \includegraphics[width=\linewidth]{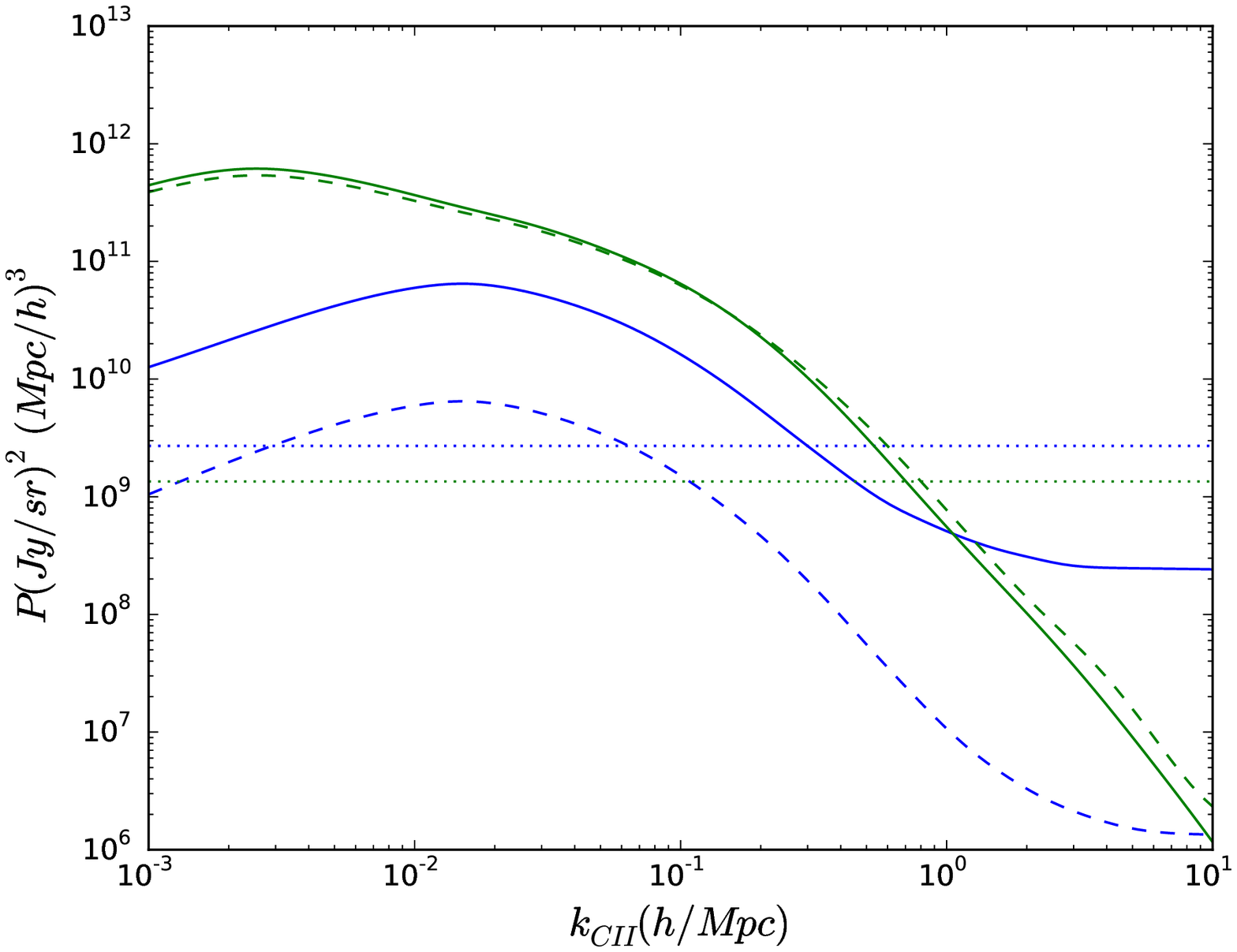}
  \caption{\label{F:1DPS_mu}\textbf{Top:} Projected [CII] power spectra with different $\mu_{CII}$ values in the fiducial model. The dotted lines are one-halo terms, the dash lines are two-halo terms, and the solid lines are the projected total power spectrum ($P^{prj}_{1h}+P^{prj}_{2h}$). \textbf{Middle:} Projected CO power spectra for different $\mu_{CII}$ and the corresponding $\mu_{CO}$. \textbf{Bottom:} The ``ave-prj'' power spectrum of [CII] (blue) and CO (green) in the fiducial model (solid lines). The dotted lines are the projected shot noise level of the CIB model for [CII] (blue) and CO (green), respectively. The ave-prj power spectra from S15 model are shown in dash lines.}
\end{figure}

\begin{figure}
\begin{center}
\includegraphics[width=\linewidth]{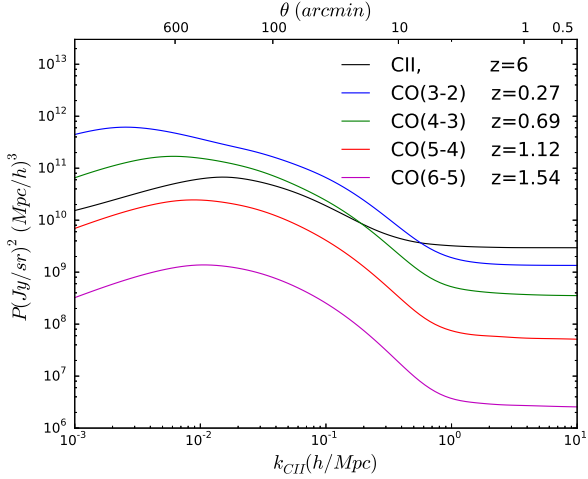}
\caption{\label{F:1DPS_mu_mult}The ``ave-prj'' power spectra of [CII] and several CO contaminants in the fiducial model. The upper axis is the corresponding observed angular scale given by $\theta=2\pi/k_{CII}\chi(z=6)$.}
\end{center}
\end{figure}

To demonstrate the deblending technique, we consider a simple case where the $z_{CII}=6$ [CII] line is blended with the brightest CO line, CO(3-2) from $z=0.27$, in an intensity mapping observing volume.  This is our fiducial model.  Fig.~\ref{F:1DPS_mu} shows the fiducial [CII] and projected CO power spectra in different $\mu_{CII}$ values. We also show power spectra obtained by averaging over $\mu_{CII}$, which we called the ``ave-prj'' power spectrum hereafter. For comparison, we plot the ave-prj power spectra using the $SFR-M$ relation in \citet{2015ApJ...806..209S} (hereafter S15 model). The ave-prj CO power spectra for other {\it J} transitions are shown in Fig.~\ref{F:1DPS_mu_mult}. 

\section{MCMC-based Parameter Inference}\label{S:MCMCinfer}
We construct the [CII] and projected CO power spectra with the CIB model described above, and use them as templates to extract [CII] and CO components in a simulated intensity mapping data cube.  Specifically, we use the Markov Chain Monte Carlo formalism to extract power spectrum parameters. 

\subsection{[CII] and CO Templates}
\begin{figure}
  \includegraphics[width=\linewidth]{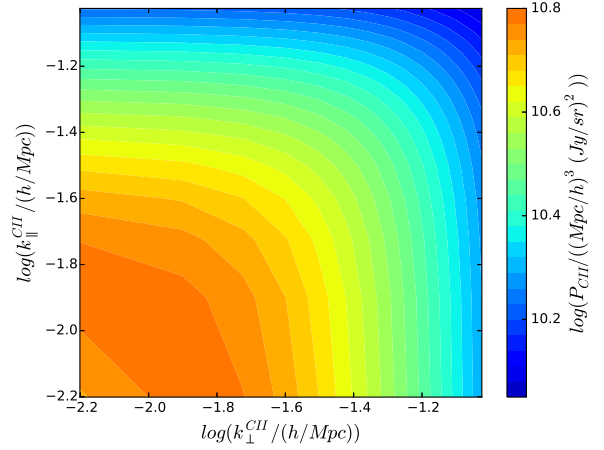}
  \includegraphics[width=\linewidth]{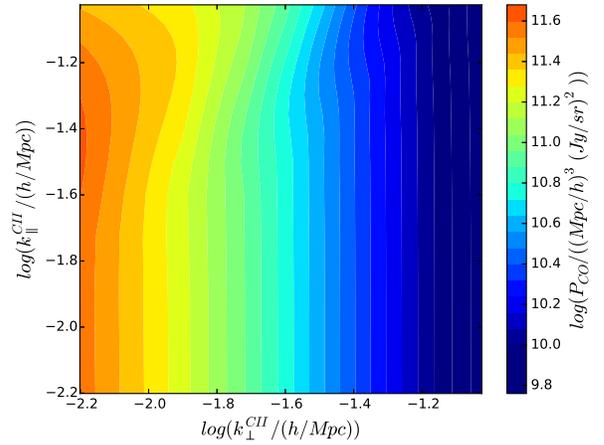}
  \caption{\label{F:Templates}[CII] (top) and CO (bottom) power spectrum templates with fiducial model in the k range specified in Sec.~\ref{S:krange}.}
\end{figure}

In summary, the [CII] and CO power spectra are derived based on Eqns. (\ref{E:rsd_1h}),~(\ref{E:rsd_2h}),~(\ref{E:rsd_1h_co}), and~(\ref{E:rsd_2h_co}) ablove.   We write the 3D power spectrum in the following form:
\begin{align}\label{E:Pline}
P_{line}(k_{CII},\mu_{CII})=(1+\frac{f(z_{line})}{\bar{b}_{line}}\mu_{line}^2)^2\nonumber\\
\times A_{line}T_{line}(k_{CII},\mu_{CII}),
\end{align}
where $A_{line}$ is a constant amplitude of the power spectrum, and $\bar{b}_{line}$ the luminosity-weighted bias of the emission line.  $T(k^\perp, k^\parallel)$ is the power spectrum template, given by the sum of the one-halo and two-halo terms in Eq.~(\ref{E:rsd_1h}),~(\ref{E:rsd_2h}),~(\ref{E:rsd_1h_co}), and~(\ref{E:rsd_2h_co}) (without the Kaiser RSD term). Fig.~\ref{F:Templates} shows the [CII] and CO power spectrum templates $T$ for the fiducial model.

We have assumed that the shape of the power spectrum templates $T$ is independent of the luminosity-halo mass relation, so that only an amplitude parameter $A$ is required to describe variations in the model.  This assumption is valid on large scales, where the bias parameter is assumed to be scale-independent and only the linear Kaiser RSD effect is important. The large-scale power spectrum is proportional to $\bar{b}^2\bar{I}^2$, where $\bar{b}$ and $\bar{I}$ are the luminosity-weighted bias and mean intensity, respectively, and reduces to the linear form: $P(k, \mu) = (1+\frac{f(z)}{\bar{b}}\mu^2)^2\bar{b}^2\bar{I}^2P_{lin}(k).$ Thus the power spectrum can be easily described by $\bar{b}$ in the Kaiser term (where we assume $f$ is fixed by our chosen cosmology) and an overall amplitude $A$ that is proportional to $\bar{b}^2\bar{I}^2$. These are the two parameters we fit with MCMC.  For this purpose, we restrict ourselves to large scales only. The exact k-space range we use is described in Sec.~\ref{S:krange}.  Below we work interchangeably in the (k , $\mu$) and ($k^\perp$, $k^\parallel$) space, where $k^\parallel$ is the Fourier wavenumber parallel to the light-of-sight and $k^\perp$ perpendicular to it, and $k=\sqrt{ (k^\perp)^2 + (k^\parallel) ^2}$ and $\mu = k^\parallel/k$.

\subsection{k-space Range}\label{S:krange}

We consider a cubic survey size with a spatial dimension of $10 \times 10$ deg$^2$. At $z=6$, this  corresponds to a linear comoving scale of $952\ Mpc/h$;  assuming the same comoving scale in the light-of-sight direction, we have an intensity mapping survey volume of $(952)^3  (Mpc/h)^3$.  We restrict our analysis to large scales only, with $k_{CII} < 0.1 h/Mpc$.  For [CII]  at $z=6$, the smallest accessible $k^\perp$ mode is $k^{min}_{CII}=6.3\times 10^{-3}\ h/Mpc$, and we choose the maximum to be $k^{max}_{CII}=9.5\times 10^{-2}\ h/Mpc$. We consider the same limit for $k^\parallel$. Thus we have $15 \times 15$ k space pixels in the mock data and the templates.

\subsection{Mock Observed Power Spectrum Construction} 
We generate a mock data power spectrum $P_{data}$, which consists of the redshifted [CII] and CO power spectra, and a noise contribution $\delta P$:
\begin{align}\label{E:Pdata}
P_{data}&(k_{CII},\mu_{CII})=P_{CII}(k_{CII},\mu_{CII})\nonumber\\
&+P_{CO}(k_{CII},\mu_{CII})+\delta P(k_{CII},\mu_{CII}),
\end{align}
where $P_{CII}$ and $P_{CO}$ are described in Eq.~(\ref{E:Pline}). 

$\delta P$ are random values that account for power spectrum noise. For each k-space pixel, $\delta P$ is drawn from a Gaussian distribution with zero mean and variance $\sigma_P^2(k_{CII},\mu_{CII})$:
\begin{align}\label{E:cv}
\sigma_P&(k_{CII},\mu_{CII})=\frac{1}{\sqrt{N(k_{CII},\mu_{CII})}}\nonumber\\
&\times(P_{data}(k_{CII},k_\parallel)+P_{sh}^{CII}+P_{sh}^{CO}+P_{n}),
\end{align}
where the $P_{data}$ term accounts for cosmic variance, $P_{sh}^{CII}$ and $P_{sh}^{CO}$ are the shot noise contributions (see Eq.~\ref{E:Psh}), and $P_n$ is the instrument noise power spectrum.  In this work, we assume $P_n=1.77\times 10^9\ (Jy/sr)^2(Mpc/h)^3$, which is consistent with the thermal noise level of current generation of planned [CII] intensity mapping experiments, such as the TIME-{\it pilot} \citep{2014SPIE.9153E..1WC}; although TIME-{\it pilot} plans a smaller survey volume than considered here. $N(k_{CII},\mu_{CII})$ is the number of pixels in each k bin. 

We do not include the shot noise and instrument white noise contributions in the template and in the mock data, since we assume they can be measured and subtracted before the template fitting process.  In a real experiment, these constant noises are the dominant ``signals'' at high-k, so we are able to infer the noise level from the high k modes and subtract them from the data.

\subsection{MCMC Implementation}
For a given set of parameters $\left \{  A_{CII},A_{CO},b_{CII},b_{CO}\right \}$, the model power spectrum $P_{model}$ is given by
\begin{align}\label{E:Pmodel}
&P_{model}(k_{CII},\mu_{CII})\nonumber\\
=&(1+\frac{f(z_{CII})}{\bar{b}_{CII}}\mu_{CII}^2)^2A_{CII}T_{CII}(k_{CII},\mu_{CII})
\nonumber\\
+&(1+\frac{f(z_{CO})}{\bar{b}_{CO}}\mu_{CO}^2)^2A_{CO}T_{CO}(k_{CII},\mu_{CII}).
\end{align}

The log-likelihood expression is
\begin{align}
ln\  \mathcal{L}&=-\frac{1}{2}\sum_{k} \{ \frac{(P_{data}(k_{CII},\mu_{CII})-P_{model}(k_{CII},\mu_{CII}))^2}{(\sigma_P(k_{CII},\mu_{CII}))^2}\nonumber\\
&+ln \left [ 2\pi(\sigma_P(k_{CII},\mu_{CII}))^2 \right ] \}.
\end{align}

We set flat priors for $A_{CII}$ and $A_{CO}$ in the range of $\left [ 10^{-6},10^6 \right ]$ (fiducial value =1), and flat priors for  $\bar{b}_{CII}$ and $\bar{b}_{CO}$ between $\left [ 0.1,20 \right ]$, and zero otherwise.

We use Python package \textit{emcee} v2.1.0 \citep{2013PASP..125..306F} to perform the MCMC analysis. We use an ensemble of 1000 walkers taking 1000 steps after 1000 burn-in steps.

\section{Results}\label{S:results}
\subsection{Fiducial Model}\label{S:CIB_CIB}

\begin{figure}
\begin{center}
\includegraphics[width=\linewidth]{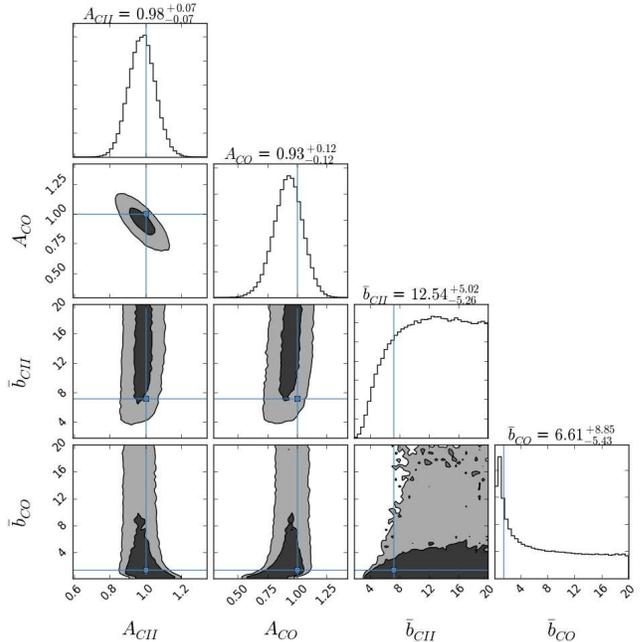}
\caption{\label{F:corner_CIB} MCMC posterior distribution of the fiducial case. The contours marked the 68\% and 95\% confidence interval in the parameter space. Crosshairs indicate the input value of the data:$\left \{ A_{CII}=1,\ A_{CO}=1,\ \bar{b}_{CII}=7.20,\ \bar{b}_{CO}=1.48\right \}$.}
\end{center}
\end{figure}

Fig.~\ref{F:corner_CIB} shows the posterior distribution of the four parameters in the fiducial case where $\left \{A_{CII},\ A_{CO},\ \bar{b}_{CII},\ \bar{b}_{CO}\right \}= \{1,\ 1,\ 7.20,\ 1.48\}$. For each point in the four-dimensional parameter space, we construct an ave-prj power spectrum by averaging over $\mu_{CII}$ of a 3D power spectrum specified by these parameters. Instead of examining the amplitude parameters $A_{CII}$ and $A_{CO}$, we use the ave-prj power spectra to compare the input and output amplitudes.

\begin{figure*}
\minipage{0.5\linewidth}
  \includegraphics[width=\linewidth]{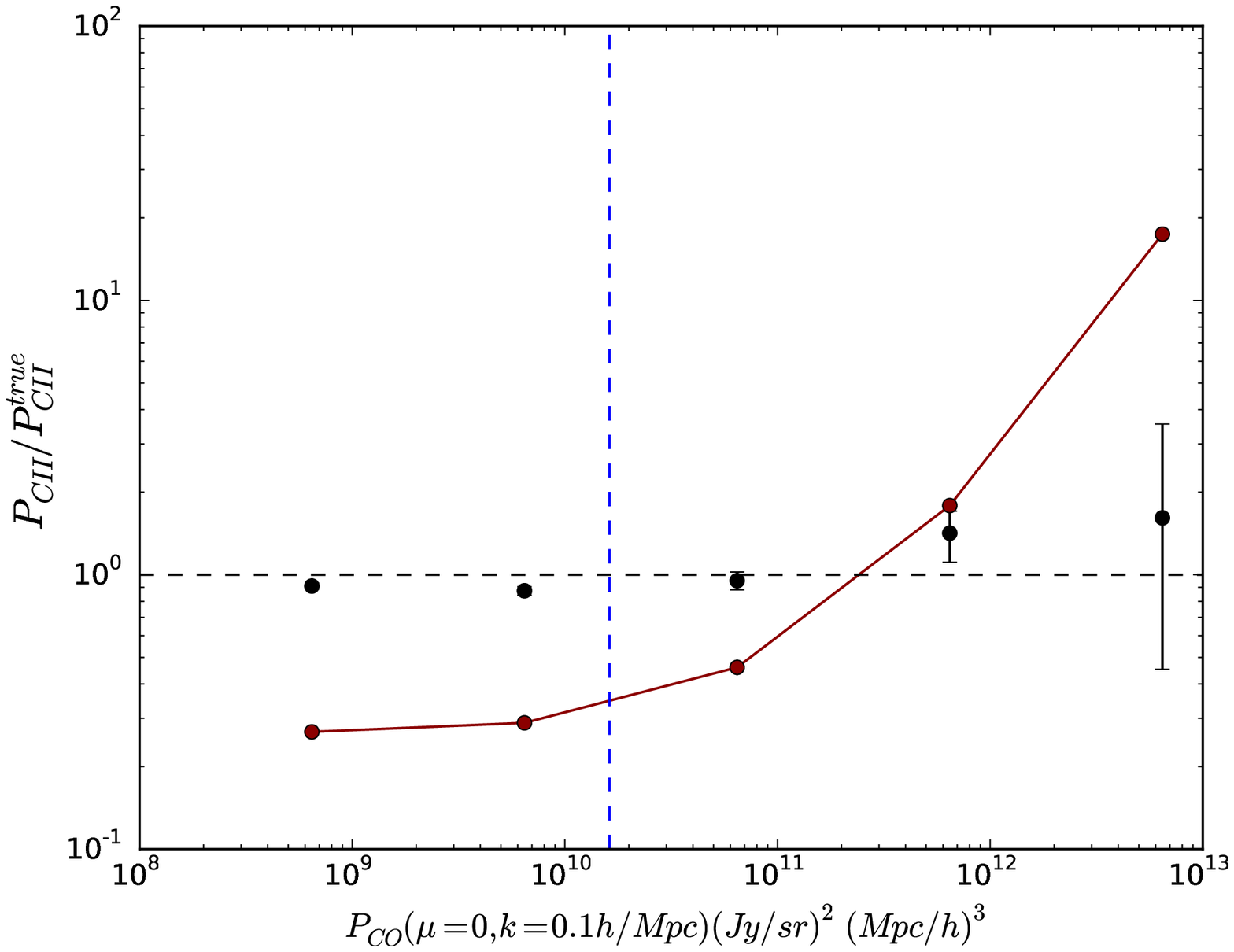}
\endminipage\hfill
\minipage{0.5\linewidth}
  \includegraphics[width=\linewidth]{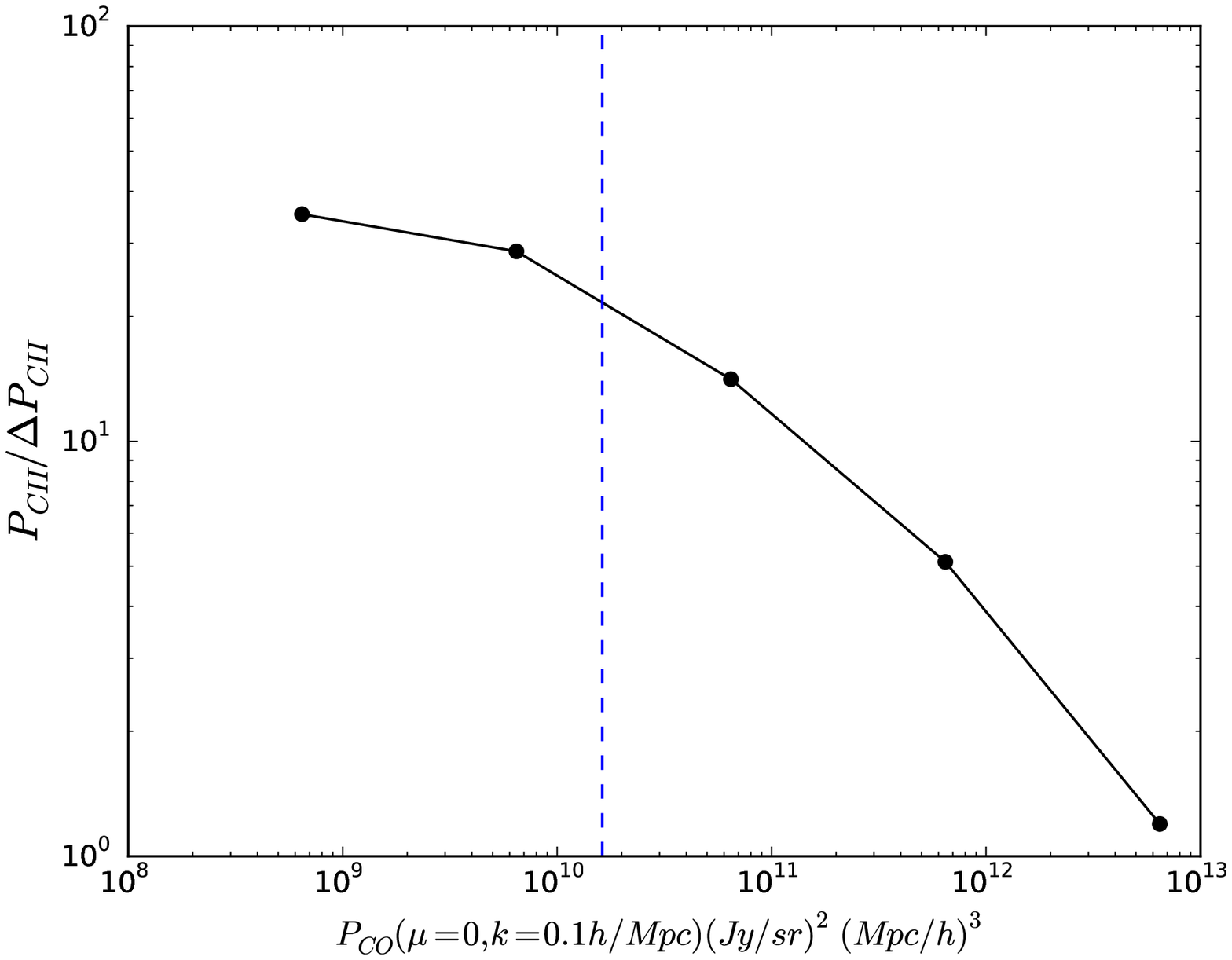}
\endminipage\hfill
  \caption{\label{F:Aco_CIB}\textbf{Left:} The ave-prj power spectrum amplitude from MCMC relative to the input value in the mock data. The inputs are $A_{CII}=1$, and $A_{CO}=[0.01,0.1,1,10,100]$ from left to right. The x-axis is expressed in the CO ave-prj power spectrum at $k_{CII}=0.1\ h/Mpc$ for better comparison. The blue dash line indicates the input [CII] prj-ave power spectrum at $k_{CII}=0.1\ h/Mpc$. The error bars are the 68\% confident interval given by MCMC. The dark red line indicates the $A_{\sigma}$ values (see text). \textbf{Right:} SNR of the ave-prj [CII] power spectrum amplitude given by MCMC.}
\end{figure*}

\begin{figure*}
\minipage{0.5\linewidth}
  \includegraphics[width=\linewidth]{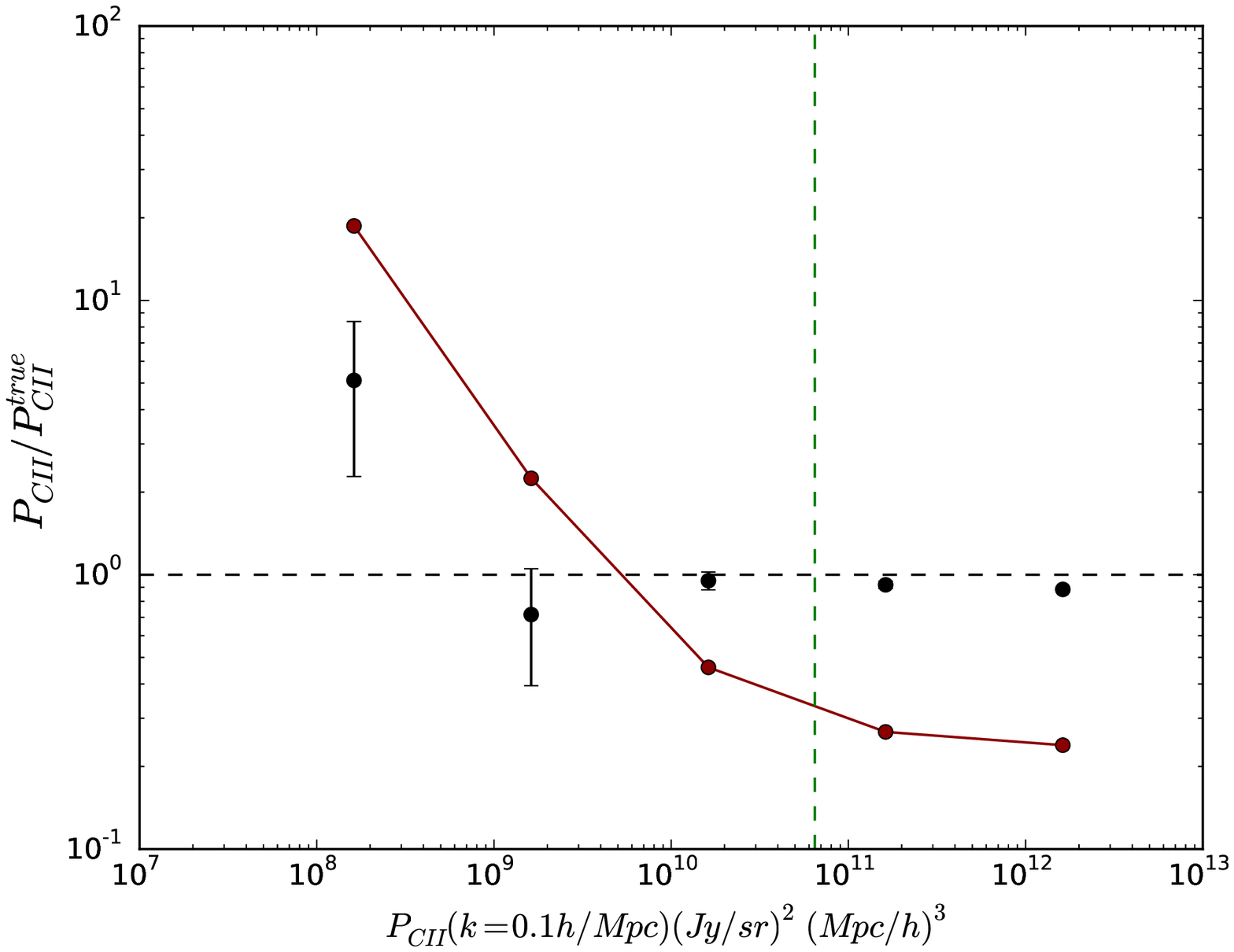}
\endminipage\hfill
\minipage{0.5\linewidth}
  \includegraphics[width=\linewidth]{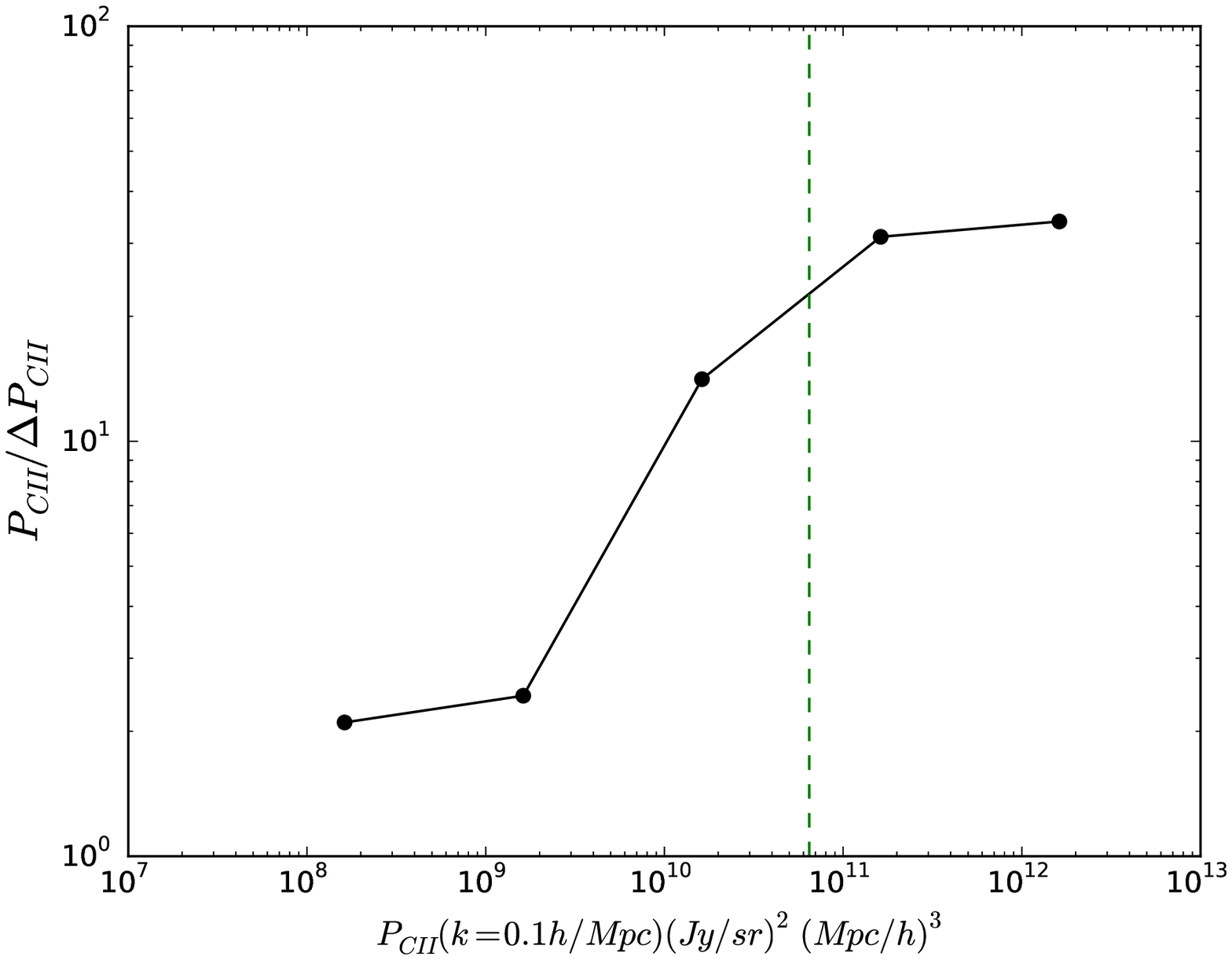}
\endminipage\hfill
  \caption{\label{F:Acii_CIB}\textbf{Left:} The ave-prj power spectrum amplitude from MCMC relative to the input value in the mock data. The inputs are $A_{CO}=1$, and $A_{CII}=[0.01,0.1,1,10,100]$ from left to right. The x-axis is expressed in the [CII] ave-prj power spectrum at $k_{CII}=0.1\ h/Mpc$ for better comparison. The green dash line indicates the input CO prj-ave power spectrum at $k_{CII}=0.1\ h/Mpc$. The error bars are the 68\% confident interval given by MCMC. The dark red line indicates the $A_{\sigma}$ values (see text). \textbf{Right:} SNR of the ave-prj [CII] power spectrum amplitude given by MCMC.}
\end{figure*}

Next, we consider more general cases by changing the input amplitudes $A_{CII}$ and $A_{CO}$ in the mock data. The [CII] and CO shot noise levels also vary accordingly with the clustering amplitudes. We first fix $A_{CII}=1$ and run MCMC with $A_{CO}=[0.01,0.1,1,10,100]$. Fig.~\ref{F:Aco_CIB} (left) shows the 68\% confidence interval of ave-prj [CII] power spectrum amplitude relative to the input, $P^{true}_{CII}$. For comparison, we also calculate the amplitude of the noise power spectrum $\sigma_P$ (see Eq.~\ref{E:cv}) relative to the best fit value of ave-prj [CII] power spectrum $P^{best}_{CII}$. We define a quantity $A_{\sigma}$ to be the median value of the ratio $\sigma_P/P^{best}_{CII}$ over the $15\times 15$ k-space pixels.  $A_{\sigma}$ serves as an indicator of the available information content level, which we discuss further in Sec.~\ref{S:detection}. The $SNR\equiv P_{CII}/\Delta P_{CII}$ of [CII] ave-prj power spectrum is shown in Fig.~\ref{F:Aco_CIB} (right), where $\Delta P_{CII}$ is the standard deviation of the ave-prj power spectrum amplitudes given by MCMC.  We then fix $A_{CO}=1$ and set $A_{CII}=[0.01,0.1,1,10,100]$ to repeat the exercise.  The results are shown in Fig.~\ref{F:Acii_CIB}. 

In the three cases with high [CII] to CO ratios, where $A_{\sigma}<1$, the mean of the ave-prj [CII] power spectrum can be reproduced within $10\%$ deviation from the true value, with $SNR>10$ in both tests.  Not all of the true ave-prj [CII] power spectrum amplitudes, however, fall in the $68\%$ interval of the MCMC distributions.  To understand this inconsistency, we run 100 realizations of mock data with the fiducial case ($A_{CII}=A_{CO}=1$).  We find that in 63 of the 100 runs, the ave-prj $P_{CII}^{true}$ does fall in the 68\% interval of the MCMC distribution.  This suggests that parameter degeneracy may exist in the fitting, so that some of the fitted amplitudes deviate slightly from the true values. This issue might be resolved by adopting tighter constraints on the input $\bar{b}_{CII}$ and $\bar{b}_{CO}$. We will investigate this degeneracy in future work.  

For the rest, where $A_{\sigma}>1$, the $SNR$ given by MCMC degrade to less than 10 and one of the mean value is biased by a factor of 4;  we find $A_{\sigma}$ to be a good indicator in determining the reliability of the fitting. 

\begin{figure*}
\begin{center}
\includegraphics[width=\linewidth]{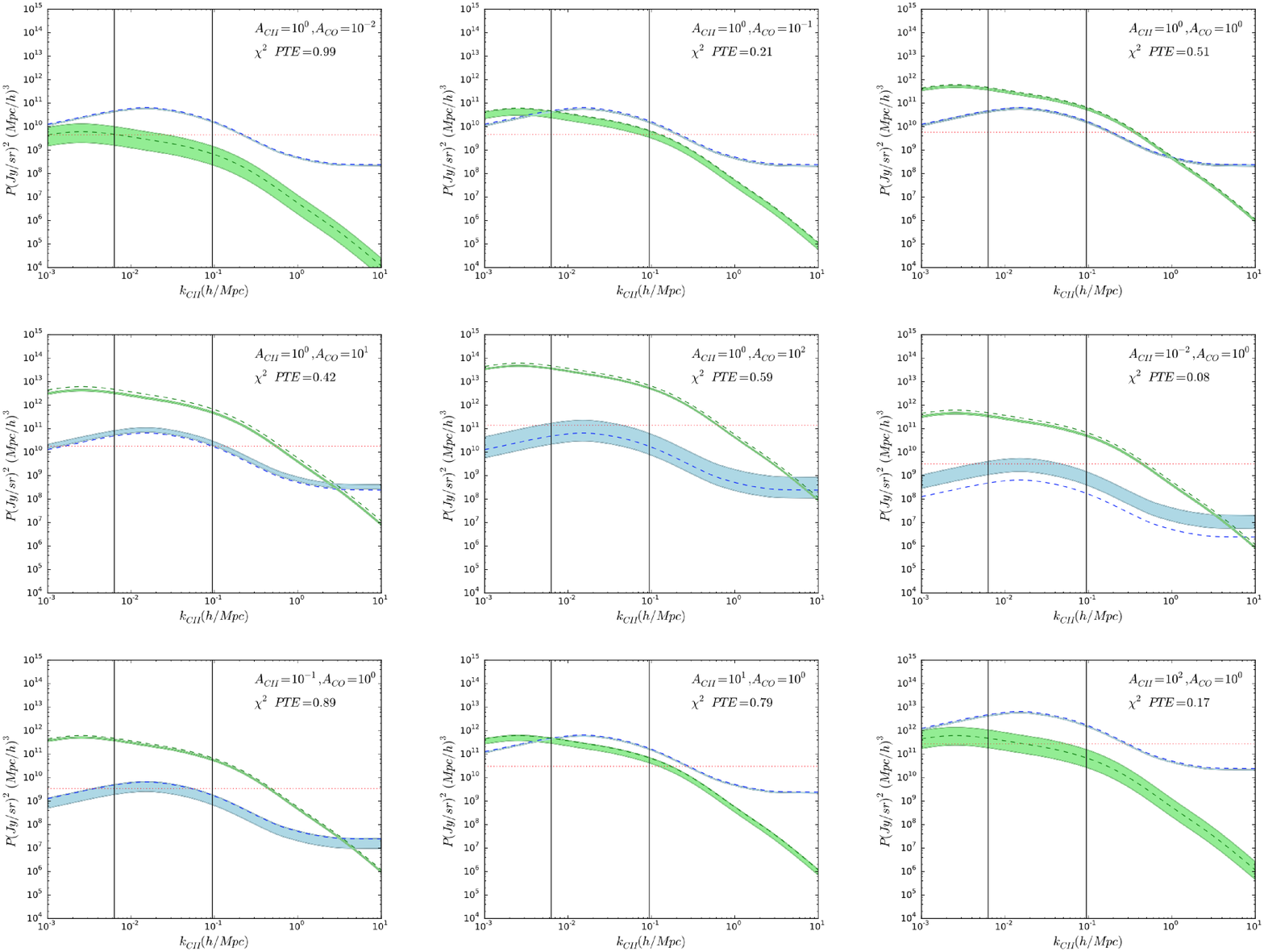}
  \caption{\label{F:9plots_CIB}The ave-prj power spectrum with different combination of input $A_{CII}$ and $A_{CO}$. The blue and green regions are the 68\% confidence interval of [CII] and CO ave-prj power spectrum respectively. The dash lines are the input [CII](blue) and CO(green) ave-prj power spectrum of the mock data. The red dot line indicates $P_{sh}^{CII}+P_{sh}^{CO}+P_n$, which is the constant power spectrum in the data that contribute to the noise in k space (see Eq.~\ref{E:cv}). The two vertical black lines are $k^{min}_{CII}$ and $k^{max}_{CII}$ marking the k space region we use in template fitting. The $\chi^2$ PTE for each MCMC fit is also provided.}
\end{center}
\end{figure*}

Fig.~\ref{F:9plots_CIB} shows the ave-prj power spectrum for all nine combinations of input $A_{CII}$ and $A_{CO}$ we discuss above. Reproducing the best-fit mock data with the four parameters $\left \{ A_{CII},\ A_{CO},\ \bar{b}_{CII},\ \bar{b}_{CO}\right \}$ and comparing with the input mock data, we calculate the $\chi^2$  in each cases, and the probability to exceed (PTE) the observed value if MCMC gives correct parameters. The PTE values are also shown in Fig.~\ref{F:9plots_CIB}.

\subsection{SNR Dependence on Noise Level}\label{S:noise}
\begin{figure}
\begin{center}
\includegraphics[width=\linewidth]{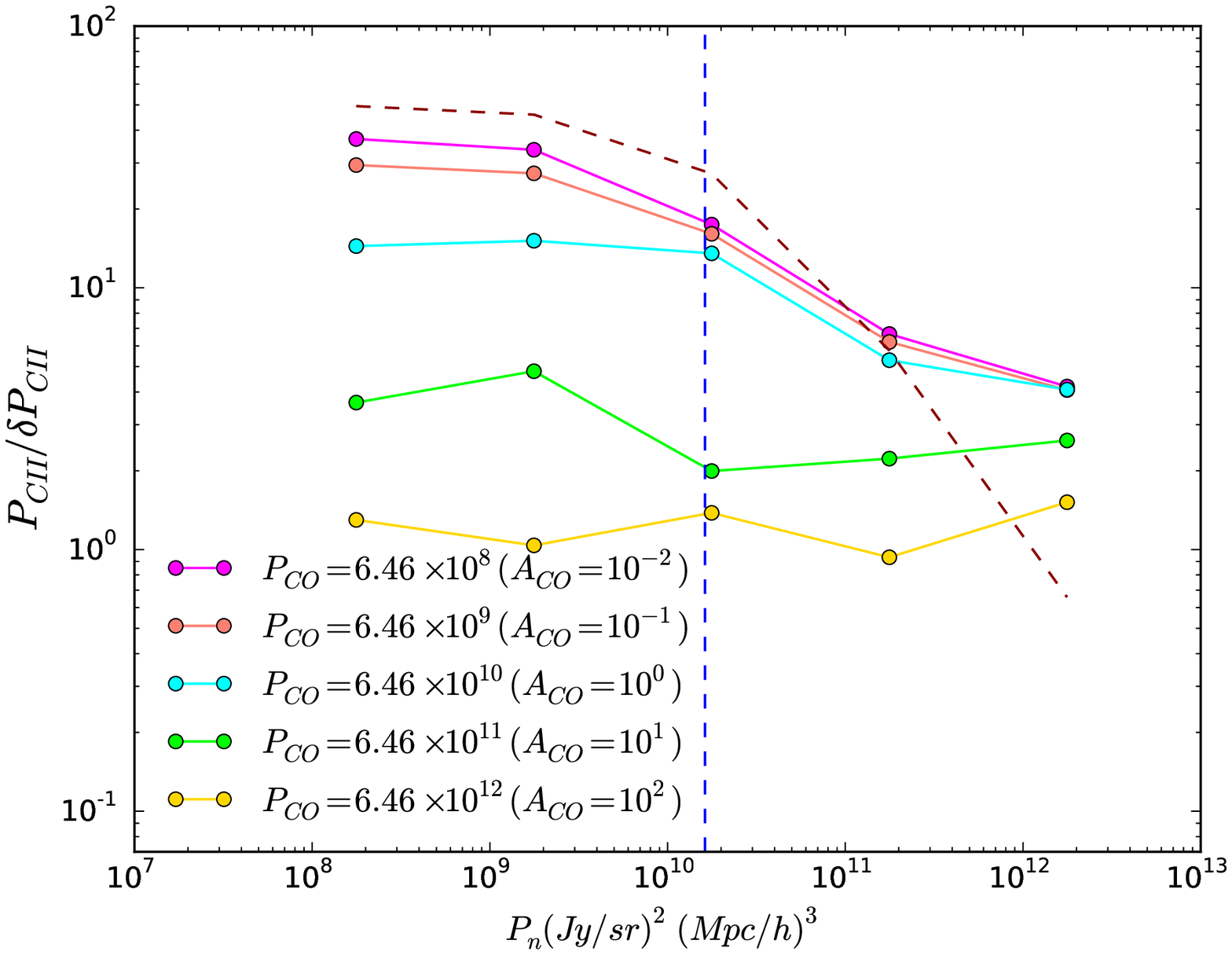}
\caption{\label{F:Pn}The SNR of the ave-prj [CII] power spectrum amplitude as a function of $P_n$ and $P_{CO}$. The $P_{CO}$ written on the top right corner is the ave-prj power spectrum at $k=0.1\ h/Mpc$. The blue dash line marks the [CII] ave-prj power spectrum at $k=0.1\ h/Mpc$. The dark-red dash line is the SNR derived from mode counting (see text).  }
\end{center}
\end{figure}

We assume the instrument noise can be subtracted from the data power spectrum before template fitting, but it still contributes a k-space noise in Eq.~(\ref{E:cv}). Here we investigate the effect of instrument noise level $P_n$ on the fitted results. We again fix input $A_{CII}=1$ and $A_{CO}=[0.01,0.1,1,10,100]$, and run a series of cases by changing $P_n$ to be $[0.1,1,10,100,1000]$ times of the initial $P_n$ value ($P_{n,i}=1.77\times 10^9\ (Jy/sr)^2(Mpc/h)^3$). The results are shown in Fig.~\ref{F:Pn}. For comparison, We also calculate the theoretical SNR on the [CII] power spectrum, which can be expressed as
\begin{equation}
SNR=\sqrt{\sum_{k}\left ( \frac{P_{CII}(k_{CII},\mu_{CII})}{\sigma_P(k_{CII},\mu_{CII})} \right )^2},
\end{equation}
where $\sigma_P$ is given by Eq.~(\ref{E:cv}), and summing over all k-space pixels.

Note that all of the highest $P_n$ cases show biased results, and some of the large $P_n$ cases also give amplitudes that deviate significantly from the input values.  This may be indicative that given the survey size we consider in this work,  $P_n\approx 10^{10}(Jy/sr)^2(Mpc/h)^3$ is the maximum allowed instrument noise for detecting the CIB-based [CII] signals. We will conduct a more detailed investigation in future work.

\subsection{Multiple Foreground Lines}\label{S:Mult_line}
So far we have only considered the brightest foreground CO (3-2) line at $z=0.27$ to the $z=6$ [CII] signal. Here we extend the technique to incorporate two more foreground lines: CO(4-3) from $z=0.69$, and CO(5-4) from $z=1.12$. The power spectra of $J>5$ transition lines are more than two orders of magnitude lower than the expected [CII] signal, and we do not consider them in this paper. 

The extra CO lines can be incorporated by extending Eq.~(\ref{E:Pdata}) and Eq.~(\ref{E:Pmodel}) with two more CO terms, which introduce two more amplitudes and bias factors in the MCMC procedure. Thus we now fit the mock data with eight parameters in the same k-space defined before.  
We set all the input amplitude to be unity, so the mock data is consistent with the CIB model prediction. 

\begin{figure*}
\includegraphics[width=\linewidth]{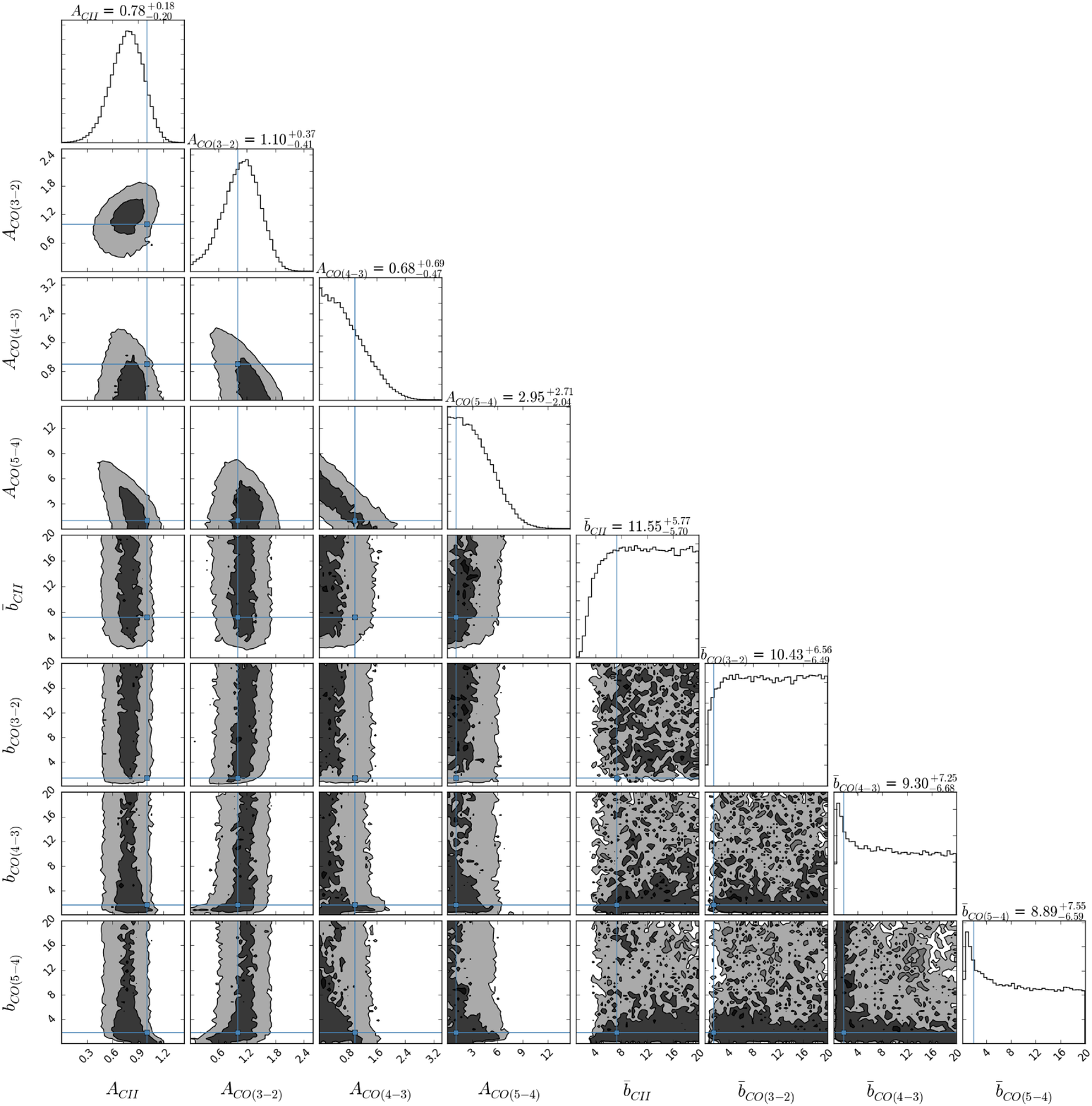}
\caption{\label{F:corner_8par} MCMC posterior distribution on the parameter space with three CO foreground lines. The contours marked the 68\% and 95\% confidence interval in the parameter space. Crosshairs indicate the input value of the data:$\left \{ A_{CII}=1,\ A_{CO(3-2)}=1,\ A_{CO(4-3)}=1,\  A_{CO(5-4)}=1,\ \bar{b}_{CII}=7.20,\ \bar{b}_{CO(3-2)}=1.48,\ \bar{b}_{CO(4-3)}=1.70,\ \bar{b}_{CO(5-4)}=1.94\right \}$.}
\end{figure*}

The result is shown in Fig.~\ref{F:corner_8par}. The [CII] ave-prj amplitude given by MCMC has a $SNR=4.12$, and the input [CII] ave-prj amplitude slightly falls outside the $68\%$ confidence interval.  The technique appears to be valid even in the presence of multiple foreground lines that overwhelm the [CII] signals in the k range we consider, although at the cost of SNR for the extracted [CII] signal.

\section{Discussion}\label{S:discussion}

\subsection{Model Dependence}
\begin{figure}
\begin{center}
\includegraphics[width=\linewidth]{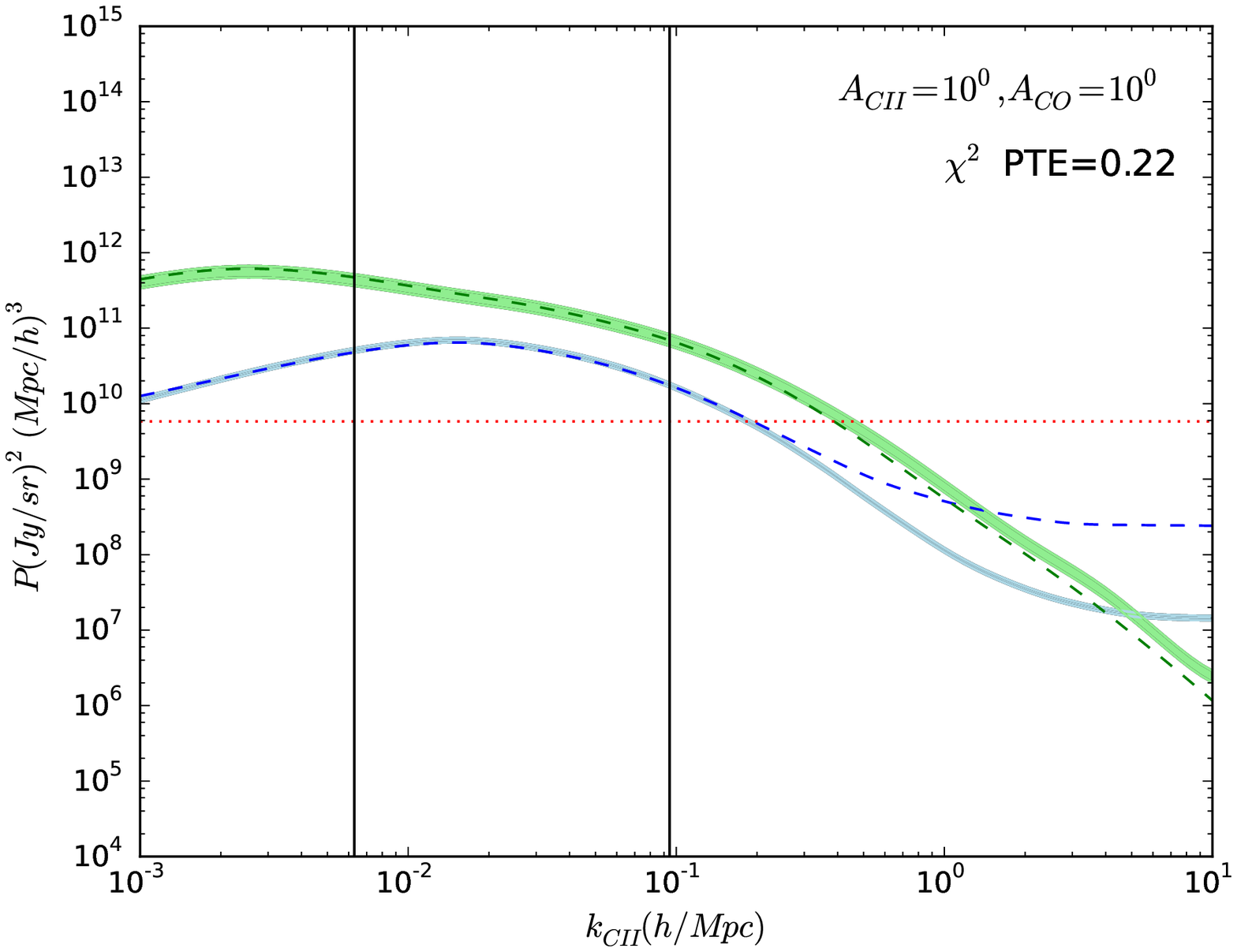}
\caption{\label{F:CS} The ave-prj power spectrum of fiducial CIB model fitted with S15 template. The blue and green regions are the 68\% confidence interval of [CII] and CO ave-prj power spectrum respectively. The dash lines are the input [CII](blue) and CO(green) ave-prj power spectrum of the mock data. The red dot line indicates $P_{sh}^{CII}+P_{sh}^{CO}+P_n$, which is the constant power spectrum in the data that contribute to the noise in k space (see Eq.~\ref{E:cv}). The two vertical black lines are $k^{min}_{CII}$ and $k^{max}_{CII}$ marking the k space region we use in template fitting. The $\chi^2$ PTE for each MCMC fit is also provided.}
\end{center}
\end{figure}

In reality, if the templates do not perfectly describe the true signal intensity field, there will be amplitude and shape discrepancies between the template and true signal power spectra. The overall amplitude discrepancy can be fully absorbed by the amplitude parameters $A_{CII}$ and $A_{CO}$ in our fitting process. Power spectrum shape difference may arise from different assumed $L-M$ relation, but since we restrict our fitting to the large-scale clustering terms, the procedure is not susceptible to incorrect model assumptions. As a sanity check, we use templates generated from the S15 model to fit the same set of mock data discussed in Sec.~\ref{S:CIB_CIB}, which are constructed with the CIB model. Fig.~\ref{F:CS} shows the ave-prj power spectrum from fitting the S15 template to the CIB model mock data. The results for all the nine scenarios considered before are also consistent with the results presented in Sec.~\ref{S:CIB_CIB}.  The template fitting technique is robust against model uncertainties as long as only large-scale information is considered.

Conversely, this model-independent property implies that the technique constrains only the overall amplitude of the power spectrum and is not sensitive to different $L-M$ relation scenarios. 

\subsection{Model Uncertainties}\label{S:model_uncertainties}
While [CII] and CO modeling uncertainties do not bias the template fitting results, they affect the quality of the fit. In Sec.~\ref{S:model}, we show that the CIB model considered in this work gives rise to a high [CII] power and a large bias factor.  Our modeling of the different CO {\it J} rotational lines also determines the relative amplitudes of interlopers and the severity of contamination.  

We have therefore conducted a series of tests with different input [CII] and CO amplitudes and varying noise levels to account for the uncertainties.  We find that if the true [CII] power is 10 times smaller than the fiducial CIB amplitude (second-left point in Fig~\ref{F:Acii_CIB}), for instance, $A_\sigma$ goes above unity, which implies a non-detection, contrary to the more optimistic fiducial case.  

\begin{figure}
\begin{center}
\includegraphics[width=\linewidth]{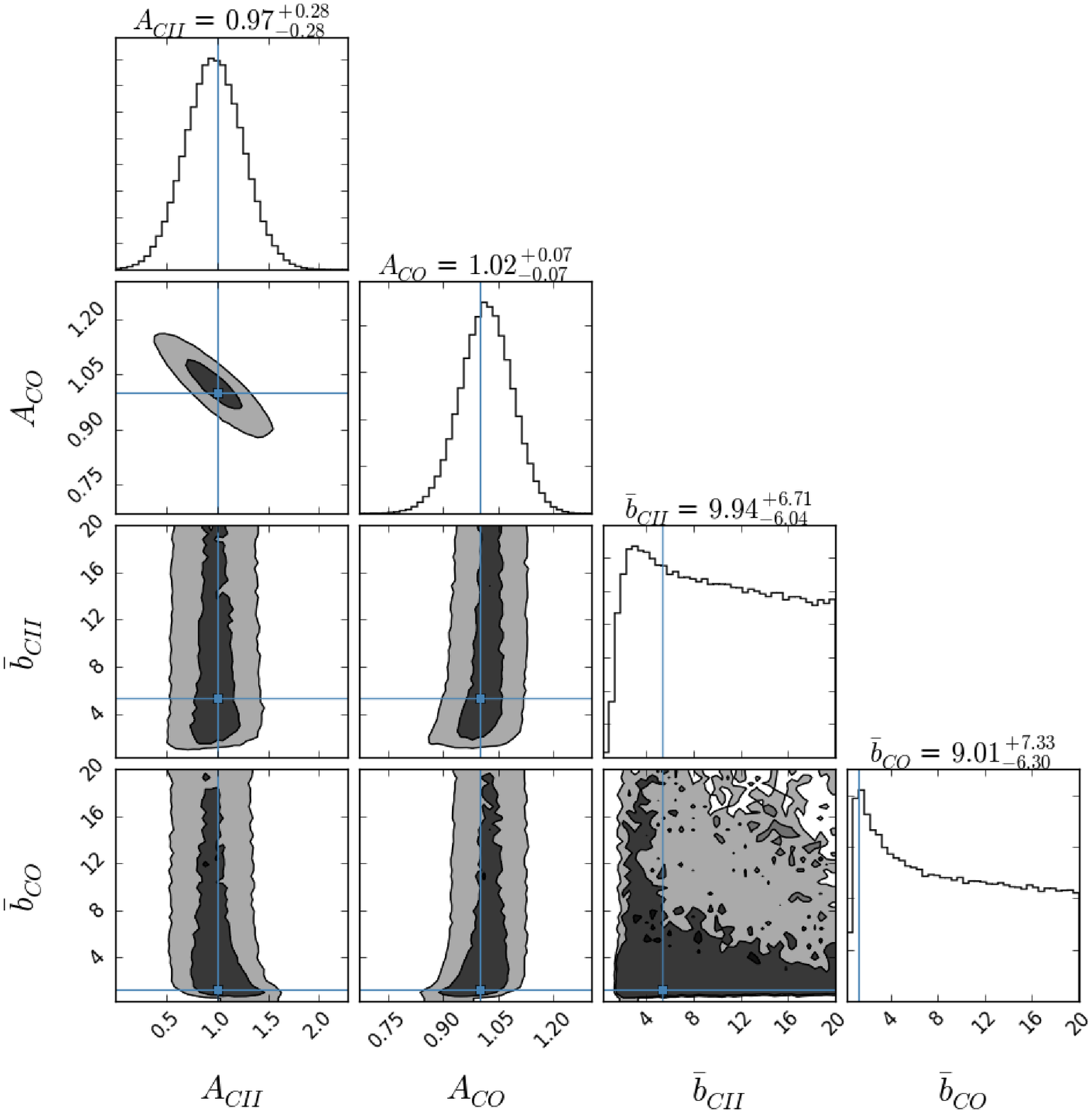}
\caption{\label{F:corner_S14} MCMC posterior distribution of the S15 case. The contours marked the 68\% and 95\% confidence interval in the parameter space. Crosshairs indicate the input value of the data:$\left \{ A_{CII}=1,\ A_{CO}=1,\ \bar{b}_{CII}=5.38,\ \bar{b}_{CO}=1.29\right \}$.}
\end{center}
\end{figure}

To get a sense of how much the results vary with the assumed models, we run simulations with mock data and templates generated from the S15 model.  Compared to the CIB model, the S15 model has slightly lower bias values for both [CII] and CO lines and a similar CO power spectrum amplitude, while the [CII] power spectrum amplitdue is lower by about one order of magnitude (see the bottom panel of Fig.~\ref{F:1DPS_mu}). Fig.~\ref{F:corner_S14} shows the MCMC result of the S15 model. The SNR the on ave-prj [CII] power spectrum in this case is 4.5.  For comparison, the fiducial CIB model has an SNR of $\approx$ 14.  We find the fitted [CII] power spectrum SNR depends sensitively on its amplitude, when overwhlemed by the CO foregrounds.

In a similar work by Lidz \& Taylor (2016, in prep), the authors modeled the power spectrum with a SFR that follows the Schechter functional form \citep{1976ApJ...203..297S}, and the [CII] power spectrum they derived at $z=7$ was about one order of magnitude smaller than our fiducial $z=6$ prediction.  Their assumed survey volume was about 5.6 times smaller than considered here.  As a result, the [CII] amplitude constraints in their study using Fisher matrix formalism appeared less optimistic than ours, but the two results are broadly in agreement.  In reality, the feasibility of this de-blending method will be highly dependent on the assumed survey geometry and signal/noise strengths. We will leave a more realistic [CII] and CO power spectrum modeling to future work, while caution that built-in detection margins are important for planned experiments.

\subsection{Continuum Foreground}
\begin{figure}
\begin{center}
\includegraphics[width=\linewidth]{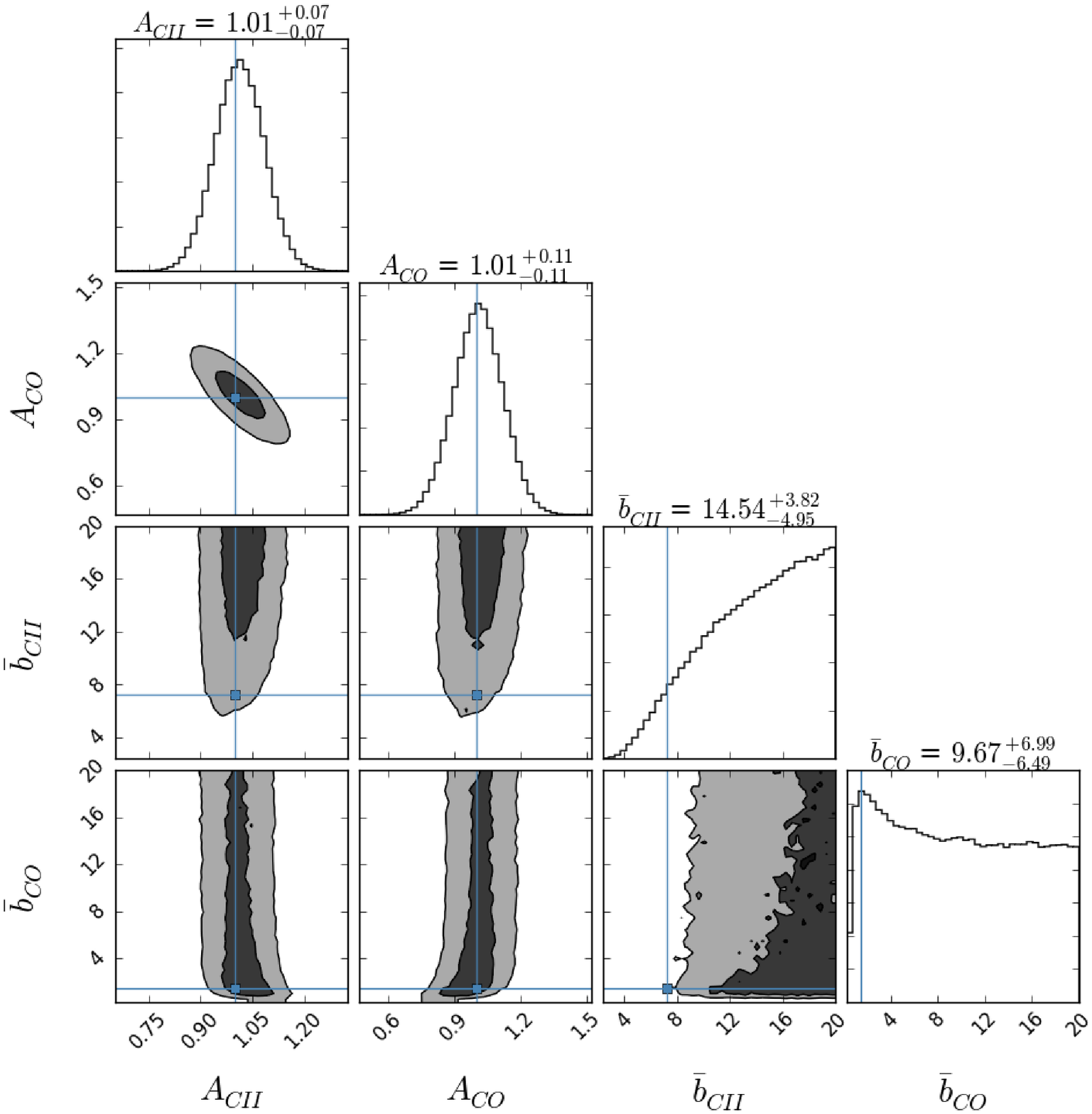}
\caption{\label{F:corner_conti} MCMC posterior distribution of the fiducial model with the lowest $k^\parallel$ mode removed.}
\end{center}
\end{figure}

In this paper we focus on the de-confusion technique that handles spectral line foregrounds.  For completeness, we note that at the frequency range of interest, $\sim 200-300$ GHz, continuum foregrounds are non-negligible and generally stronger than line foregrounds.   For our purpose, the main continuum foregrounds include the cosmic microwave background and cosmic infrared background radiations;  the two contribute comparably.  \citet{2015ApJ...806..209S} estimates that the dust continuum emission is of the order of $10^5$ Jy sr$^{-1}$, which is two or three orders of magnitude (depending on model) higher than the [CII] intensity considered in this work. However, since the continuum signals are expected to be spectrally smooth, they dominate the low $k^\parallel$ modes in power spectrum space. We therefore expect to be able to mitigate the effect by removing or avoiding the one or two lowest $k^\parallel$ modes before template fitting.  This is the same technique envoked in the well-studied field of 21-cm intensity mapping (e.g., \citet{2011PhRvD..83j3006L,2012ApJ...756..165P,2015ApJ...815...51S}) and implemented on intensity mapping data (e.g., \citet{2013MNRAS.433..639P,2013MNRAS.434L..46S,2015ApJ...814..140K,2015ApJ...809...61A}). To test the impact of continuum foreground mode removal, we run a fiducial MCMC fit, same as the case in Sec.~\ref{S:CIB_CIB} but removing the lowest $k^\parallel$ modes.  The results are shown in Fig.~\ref{F:corner_conti}. The SNR on the [CII] power spectrum remains nearly the same as the fiducial one (Fig.~\ref{F:corner_CIB}).  This simple test demonstrates that continuum foregrounds are unlikely to be a major concern, but we note that subtler issues, such as the exact number of $k^\parallel$ modes to be removed, and the amount of residual continuum in the data, need to be quantified and further tested in the future.

\subsection{Detection Limit}\label{S:detection}
In Sec.~\ref{S:CIB_CIB},  we define $A_{\sigma}$ to be the median value of $\sigma_P/P^{best}_{CII}$ and use it as an indicator of the available information content level; when $A_{\sigma} < 1$, or $\sigma_P < P^{best}_{CII}$, we recover an unbiased estimate of $P^{true}_{CII}$.  Indeed, in Sec.~\ref{S:results},
the scenarios with high CO to [CII] amplitudes result in biased extracted [CII] power spectrum amplitudes, however, the extracted amplitudes are all below the noise level (i.e. $A_{\sigma}<1$).  This suggests that the k-space noise level is a good indicator of the information content and sets the limit for extracting [CII] signals from the data, below which the MCMC likely returns biased results.  In reality, $\sigma_P$ and $P^{best}_{CII}$ are quantities that can be directly inferred from real data, and serve to evaluate the reliability of the extracted results. 

\subsection{Constraint on Bias}\label{S:Bias_constraint}
In our template fitting procedure, the bias factors are loosely constrained by MCMC (see Fig.~\ref{F:corner_CIB} and~\ref{F:corner_8par}). This can be understood by looking at the Kaiser RSD term: $(\frac{f(z))}{\bar{b}_{line}}\mu^2)^2$, where $f \approx1$, the bias is usually between 1 to 10, $\mu^2$ is a value between 0 and 1. Thus $\frac{f(z))}{\bar{b}_{line}}\mu^2$ is usually smaller than unity, sub-dominent to the overall amplitude change of the power spectrum which is absorbed in the $A$ parameter. In addition, the MCMC constraint on $\bar{b}_{CO}$ is weaker than on $\bar{b}_{CO}$. This is due to the projection effect that makes the projected $\mu_{CO}$ very small for most of the k-space pixels (see Fig.~\ref{F:1DPS_mu}), which are then not sensitive to the Kaiser effect. 

The CIB model gives a high $\bar{b}_{CII}$ value that makes the Kaiser effect too small to be detected. This can be seen from the histogram of $\bar{b}_{CII}$ given by MCMC which extend to large values that corresponds to a non-detection of Kaiser RSD effect. To test the ability to extract bias in our procedure, we run the fiducial case and setting $\bar{b}_{CII}$ to be same as the fiducial value of $\bar{b}_{CO}=1.48$. In this case, the $\bar{b}_{CII}$ can be constrained by MCMC with a SNR$>$5; while the $\bar{b}_{CO}$ is still unconstrained due to the projection effect described above. Therefore, if $\bar{b}_{CII}$ in reality is smaller than the value we considered in this work, we might be able to better constrain $\bar{b}_{CII}$. We will investigate strategies to better extract the bias information in future work.

\section{Conclusion}\label{S:conclusion}
We demonstrate the feasibility of de-blending spectral line information in the intensity mapping regime.  We consider a 3D intensity mapping survey, where multiple spectral lines at different redshifts are embedded in the same observing volume, and make use of the anisotropic shape of their respective power spectra when projected into a common comoving coordinate to disentangle the information.  We consider deblending high-redshift [CII] signals from the brighter, lower-redshift CO interlopers.  We use the halo model and cosmic infrared background measurement to construct expected CO and [CII] templates across redshifts, and use the MCMC formalism to constrain power spectrum parameters.  We show that this technique can reproduce the linear [CII] and CO power spectrum amplitudes, although with reduced signal-to-noise, given a range of CO signal strengths and noise level.  We establish an indicator to evaluate whether the fitted parameters are unbiased.  Finally, we demonstrate the ability of extracting [CII] in the presence of multiple, stronger CO foreground lines. The technique can be extended to other line blending problems to extract information of both signal and interlopers in an intensity mapping experiment. 

\acknowledgments
We are grateful to the Time-{\it Pilot} collaboration for useful inputs throughout this work. 
We thank Phil Bull, Olivier Dor{\'e}, Tony Li, Adam Lidz, Roland de Putter, Paolo Serra, Chun-Hao To, and Heidi Hao-Yi Wu for helpful discussions and valuable comments on the manuscript. 
Y-T. C. and T.-C. C. were supported in part by MoST grant 103-2112-M-001-002-MY3. T.-C. C. gratefully acknowledges the hospitality of the Caltech OBSCOS group and the Jet Propulsion Laboratory, where part of this work was carried out.

\begin{appendices}
\section{Power Spectrum Modeling}
\subsection{$L_{IR}-M$ Relation: CIB Model}\label{A:M_SFR}
In \citet{2014A&A...571A..30P}, the authors modeled the CIB emission as measured by \textit{Planck}, and parametrized the CIB specific luminosity $L_\nu$ with the following three components:

\begin{enumerate}
\item A normalized spectral energy distribution (SED), $\Theta(\nu,z)$,  for all the galaxies:
\begin{align}
&\Theta(\nu,z)\propto \nu^\beta B_\nu(T_d(z))&;\nu<\nu_0\nonumber\\
&\Theta(\nu,z)\propto \nu^{-\gamma}&;\nu\geq \nu_0,
\end{align}
where $B_\nu $ is the Planck function, and $T_d$ the redshift-dependent dust temperature:
\begin{equation}
T_d=T_0(1+z)^\alpha .
\end{equation}

\item $L-M$ relation:
They assumed the CIB luminosity is a log-normal function $\Sigma$ of halo mass $M$ with peak mass $M_{eff}$ and variance $\sigma^2_{L/M}$, 
\begin{equation}
\Sigma (M)=M\frac{1}{(2\pi \sigma^2_{L/M})^{1/2}} e^{-(log_{10}(M)-log_{10}(M_{eff}))^2/2\sigma^2_{L/M}}.
\end{equation}

\item Redshift evolution of the $L-M$ relation: 
The global normalization is parametrized by 
\begin{equation}
\Phi(z)=(1+z)^\delta.
\end{equation}

The $L-M$ ratio is assumed to increase with redshift (i.e. $\delta>0$). 
\end{enumerate}

Combining these three components, the $L_\nu-M$ relation can be written as
\begin{equation}\label{E:LCIB}
L_{\nu}(M,z)=L_0\Phi(z)\Sigma(M)\Theta(\nu,z),
\end{equation}
with an overall normalization factor $L_0$.

With the \textit{Planck} data, the authors constrained the model parameters as listed in table 9 of \citet{2014A&A...571A..30P}. Here we adopt their best fit values: \{$\alpha=0.36,\ \beta=1.75,\ \gamma=1.7,\ \delta=3.6,\ T_0=24.4\ K,\ M_{eff}=10^{12.6}M_\odot,\ \sigma^2_{L/M}=0.5,\ L_0=0.02L_\odot$\}. We integrate Eq.~(\ref{E:LCIB}) over the wavelength range of $8-1000\ \mu m$ to obtain the total infrared luminosity $L_{IR}$.

We can convert $L_{IR}$ to SFR following \citet{1998ApJ...498..541K}:
\begin{equation} \label{E:Kennicutt}
SFR/L_{IR}=1.7\times 10^{-10} M_\odot yr^{-1}L_\odot ^{-1}.
\end{equation} 

As a sanity check, we calculate the resulting star formation rate density (SFRD) by integrating the SFR over halo mass,
\begin{equation}
SFRD(z)=\int_{M_{min}}^{M_{max}}dM\frac{dN}{dM}SFR(M,z),
\end{equation}
where $dN/dM$ is the halo mass function \citep{1999MNRAS.308..119S}, and we take 
$M_{min}=10^8\ M_\odot/h$, $M_{max}=10^{15}\ M_\odot/h$. We use this integration range for all halo mass integration throughout this work.  

For comparison, we calculate the SFRD from the $SFR-M$ relation given by \citet{2011ApJ...741...70L}, \citet{2013ApJ...768...15P}, and \citet{2015ApJ...806..209S}. We also consider the SFRD given by \citet{2014ARA&A..52..415M} and \citet{2015ApJ...802L..19R}, where the SFRD is modeled by the following four-parameter functional form
\begin{equation}
SFRD(z)=a\frac{(1+z)^{b}}{1+[(1+z)/c]^{d}}\left(M_\odot yr^{-1} Mpc^{-3}\right).
\end{equation}
\citet{2014ARA&A..52..415M} used the UV and IR galaxy counts and obtained the parameters $\{ a,b,c,d\}=\{0.015,2.7,2.9,5.6\}$; while \citet{2015ApJ...802L..19R} used the joint constraint of galaxy counts and the CMB optical depth $\tau=0.066\pm 0.12$ from \citet{2015arXiv150201589P} and obtain $\{ a,b,c,d\}=\{0.01376,3.26,2.59,5.68\}$. 

\begin{figure}
\begin{center}
\includegraphics[width=\linewidth]{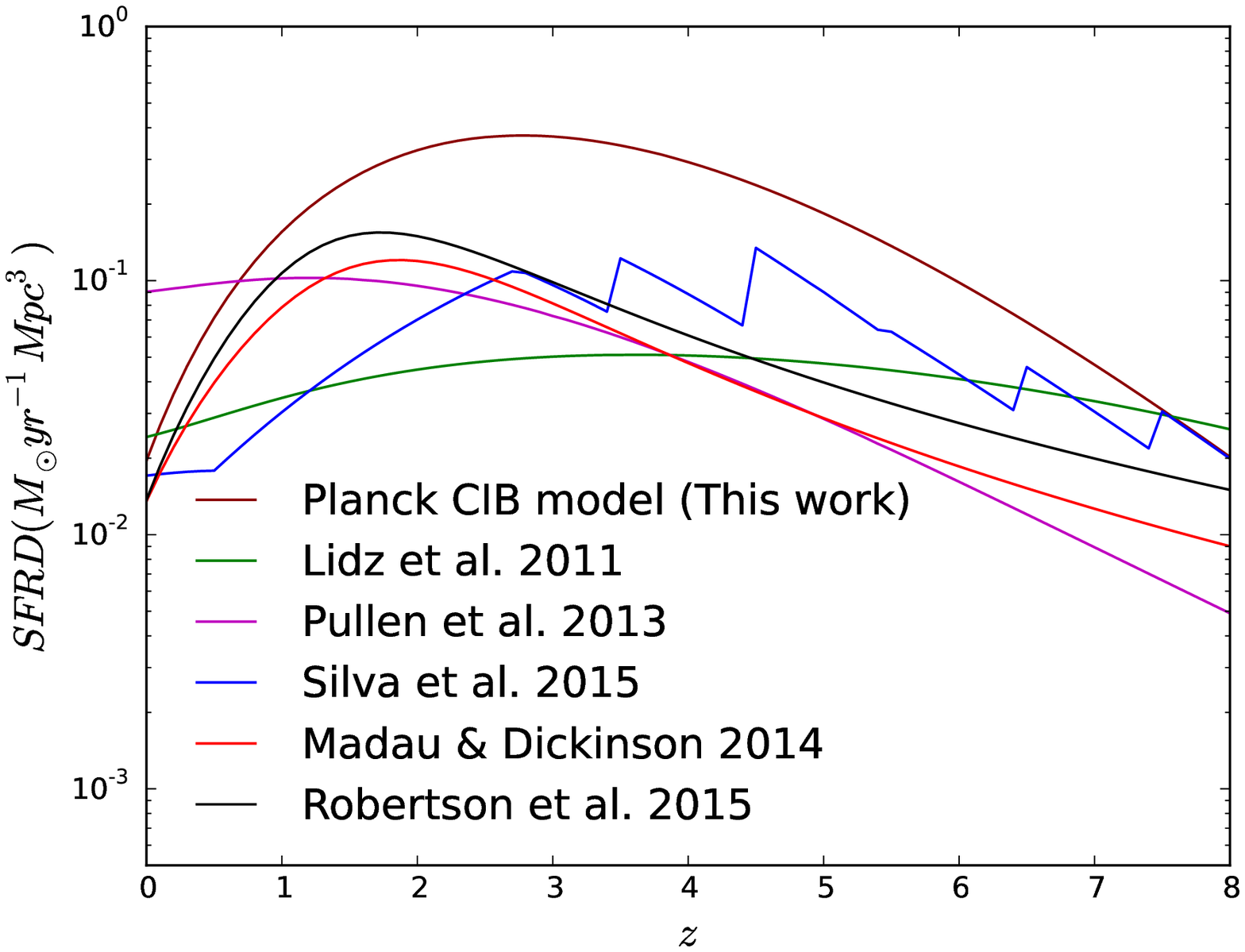}
\caption{\label{F:SFRD} The SFRD of Planck CIB model (this work) comparing with literature \citep{2011ApJ...741...70L, 2013ApJ...768...15P, 2015ApJ...806..209S, 2014ARA&A..52..415M, 2015ApJ...802L..19R}.}
\end{center}
\end{figure}

The $SFRD(z)$ of the aforementioned models are plotted in Fig.~\ref{F:SFRD}. \citet{2011ApJ...741...70L} and \citet{2013ApJ...768...15P} use a simple scaling relation to model the $SFR-M$ relation, and do not reproduce the SFRD peak at $z\sim 2-3$.  In \citet{2015ApJ...806..209S}, the $SFR-M$ relation is fitted in several redshift bins with mock galaxy catalogs from \citet{2007MNRAS.375....2D} and \citet{2011MNRAS.413..101G}. The discontinuous features in the SFRD curve are the boundaries of the redshift bins. They predict a SFRD peak at a higher redshift and the model does not agree well with galaxy counts at $z\sim 0-2$.  The SFRD from the CIB model is broadly consistent in shape with that of galaxy counts, but the CIB model predicts a systematically higher amplitude, especially at high redshifts. This results in a higher [CII] power spectrum amplitude. Furthermore, the steep $SFR-M$ relation in the CIB model also results in a high bias factor for [CII].  We discuss the implications in Sec.~\ref{S:model_uncertainties}. 

\subsection{$L_{CII}-SFR$ and $L_{CO}-SFR$ Relation}\label{A:LIR_Lline}
To connect $L_{CII}$ to SFR, we adopt the following relation based on observations of nearby late type galaxies \citep{2002A&A...385..454B}: 
\begin{equation}
L_{CII}(M,z)=1.59\times 10^{-3}L_{IR},
\end{equation}
and use Eq.~(\ref{E:Kennicutt}) to convert $L_{IR}$ to $SFR$.

For CO luminosity, we use the empirical relation from \citet{2010ApJ...714..699W}:
\begin{equation}\label{E:Lco}
L_{CO(1-0)}=3.2\times 10^4L_\odot (\frac{SFR}{M_\odot yr^{-1}})^{3/5}.
\end{equation}
For higher-$J$ rotational transitions, we assume a constant ratio between $L_{CO(1-0)}$ and $L_{CO(J-(J-1))}$, using Eq. (16) of  \citet{2009ApJ...702.1321O} and assuming an excitation temperature of $T_e=17\ K$ for all the galaxies. 

\end{appendices}
\bibliography{reference}

\end{document}